\documentclass[sigconf, screen,nonacm]{acmart}

\usepackage{longtable}
\usepackage{capt-of}
\usepackage{balance,bm}
\usepackage{color, colortbl, xcolor}
\usepackage{todonotes}
\definecolor{LightGray}{gray}{0.97}
\usepackage{pgfplots}
\usepackage{xcolor}
\usepackage{url}

\usepackage{subcaption}
\usepackage{booktabs} 
\usepackage{graphicx}
\usepackage{textcomp}
\usepackage{color,soul}
\usepackage{bm}
\usepackage{multirow}
\usepackage{wrapfig}
\usepackage{balance}
\usepackage{graphicx}  
\usepackage{enumitem}

\usepackage{colortbl}
\usepackage{arydshln}
\usepackage [english]{babel}
\usepackage [autostyle, english = american]{csquotes}
\definecolor{linkColor}{RGB}{6,125,233}
\definecolor{green}{rgb}{0.0, 0.65, 0.31}
\definecolor{bleudefrance}{rgb}{0.19, 0.55, 0.91}
\definecolor{ceruleanblue}{rgb}{0.16, 0.32, 0.75}
\definecolor{grey}{HTML}{969696}
\definecolor{violet}{HTML}{756bb1}
\definecolor{dgrey}{HTML}{01665e}
\definecolor{lgrey}{HTML}{5ab4ac}
\definecolor{dgreen}{HTML}{005a32}
\definecolor{purple}{HTML}{ae017e}

\definecolor{editCol}{HTML}{000000}
\definecolor{maskCol}{HTML}{c51b7d}
\definecolor{lrColor}{HTML}{8856a7}
\definecolor{trColor}{HTML}{d01c8b}
\definecolor{ctColor}{HTML}{4dac26}
\definecolor{brickred}{HTML}{f03b20}
\definecolor{DarkBlue}{HTML}{00008B}
\definecolor{mscolor}{HTML}{01665e}
\definecolor{nmscolor}{HTML}{bf812d}
\definecolor{lgreen}{HTML}{ccece6}
\definecolor{dolive}{HTML}{308014}

\definecolor{editCol}{HTML}{000000}
\definecolor{maskCol}{HTML}{c51b7d}
\definecolor{lrColor}{HTML}{8856a7}
\definecolor{trColor}{HTML}{d01c8b}
\definecolor{ctColor}{HTML}{4dac26}
\definecolor{brickred}{HTML}{f03b20}
\definecolor{lgreen}{HTML}{e0f3db}
\definecolor{dpink}{HTML}{CD1076}
\definecolor{pink}{HTML}{FED2D2}
\definecolor{soothinggreen}{HTML}{4dac26}
\definecolor{darkred}{HTML}{8B0000}

\definecolor{dblue}{HTML}{215F9A}
\definecolor{violet}{HTML}{8A2BE2}
\definecolor{mscolor}{HTML}{01665e}
\definecolor{nmscolor}{HTML}{d8b365}
\definecolor{deepgrey}{HTML}{525252}
\definecolor{dslate}{HTML}{2F4F4F}
\definecolor{dolive}{HTML}{556B2F}
\definecolor{teal}{HTML}{388E8E}
\definecolor{mscolor}{HTML}{01665e}
\definecolor{nmscolor}{HTML}{d8b365}

\definecolor{aicolor}{HTML}{018571}
\definecolor{occolor}{HTML}{ff7799}

\definecolor{srcolor}{HTML}{e34a33}
\definecolor{smcolor}{HTML}{253494}
\definecolor{srsmcolor}{HTML}{7fcdbb}
\definecolor{bothcolor}{HTML}{fe9929}
\definecolor{onecolor}{HTML}{018571}
\definecolor{marroon}{HTML}{881c1c}

\usepackage{mathtools}

\usepackage{amsmath}
\usepackage{array}
\usepackage{xcolor}
\usepackage{arydshln}
\usepackage{siunitx}
\colorlet{tablerowcolor4}{gray!50} 

\definecolor{improveCol}{HTML}{7b3294}
\definecolor{worsenCol}{HTML}{008837}

\definecolor{entrycol}{HTML}{CD950C}
\definecolor{exitcol}{HTML}{003057}

\newcommand*{\textlabel}[2]{%
  \edef\@currentlabel{#1}
  \phantomsection
  #1\label{#2}
}

\colorlet{tableheadcolor}{gray!25} 
\colorlet{tablerowcolor}{gray!15} 
\colorlet{tablerowcolor2}{gray!45} 
\colorlet{tablerowcolor3}{gray!25} 

\newcommand{\rowcollight}{\rowcolor{LightGray}} %
\usepackage{capt-of}
\usepackage{balance,bm}
\usepackage{color, colortbl, xcolor}

\usepackage{booktabs}  
\usepackage{url}
\usepackage{hyperref}
\hypersetup{
    colorlinks=true,
}

\usepackage{colortbl}
\usepackage{arydshln}
\usepackage [english]{babel}
\usepackage [autostyle, english = american]{csquotes}
\definecolor{linkColor}{RGB}{6,125,233}
\definecolor{green}{rgb}{0.0, 0.65, 0.31}
\definecolor{bleudefrance}{rgb}{0.19, 0.55, 0.91}
\definecolor{ceruleanblue}{rgb}{0.16, 0.32, 0.75}
\definecolor{grey}{HTML}{969696}
\definecolor{violet}{HTML}{756bb1}
\definecolor{dgrey}{HTML}{01665e}
\definecolor{lgrey}{HTML}{5ab4ac}
\definecolor{dgreen}{HTML}{005a32}
\definecolor{purple}{HTML}{ae017e}

\definecolor{editCol}{HTML}{000000}
\definecolor{maskCol}{HTML}{c51b7d}
\definecolor{lrColor}{HTML}{8856a7}
\definecolor{trColor}{HTML}{d01c8b}
\definecolor{ctColor}{HTML}{4dac26}
\definecolor{brickred}{HTML}{f03b20}
\definecolor{DarkBlue}{HTML}{00008B}
\definecolor{mscolor}{HTML}{01665e}
\definecolor{nmscolor}{HTML}{bf812d}
\definecolor{lgreen}{HTML}{ccece6}
\definecolor{dolive}{HTML}{308014}

\definecolor{editCol}{HTML}{000000}
\definecolor{maskCol}{HTML}{c51b7d}
\definecolor{lrColor}{HTML}{8856a7}
\definecolor{trColor}{HTML}{d01c8b}
\definecolor{ctColor}{HTML}{4dac26}
\definecolor{brickred}{HTML}{f03b20}
\definecolor{lgreen}{HTML}{e0f3db}
\definecolor{dpink}{HTML}{CD1076}
\definecolor{pink}{HTML}{FED2D2}
\definecolor{soothinggreen}{HTML}{4dac26}

\definecolor{dblue}{HTML}{104E8B}
\definecolor{violet}{HTML}{8A2BE2}
\definecolor{mscolor}{HTML}{01665e}
\definecolor{nmscolor}{HTML}{d8b365}
\definecolor{deepgrey}{HTML}{525252}
\definecolor{dslate}{HTML}{2F4F4F}
\definecolor{dolive}{HTML}{556B2F}
\definecolor{teal}{HTML}{388E8E}
\definecolor{mscolor}{HTML}{01665e}
\definecolor{nmscolor}{HTML}{d8b365}

\definecolor{aicolor}{HTML}{018571}
\definecolor{occolor}{HTML}{ff7799}

\definecolor{srcolor}{HTML}{e34a33}
\definecolor{smcolor}{HTML}{253494}
\definecolor{srsmcolor}{HTML}{7fcdbb}
\definecolor{bothcolor}{HTML}{fe9929}
\definecolor{onecolor}{HTML}{018571}
\definecolor{marroon}{HTML}{881c1c}

\usepackage{mathtools}

\usepackage{amsmath}
\usepackage{array}
\usepackage{xcolor}
\usepackage{arydshln}
\usepackage{siunitx}
\colorlet{tablerowcolor4}{gray!50} 

\usepackage{tcolorbox}

\colorlet{tableheadcolor}{gray!25} 
\colorlet{tablerowcolor}{gray!15} 
\colorlet{tablerowcolor2}{gray!45} 
\colorlet{tablerowcolor3}{gray!25} 

\newif{\ifhidecomments}
  \hidecommentsfalse 
\ifhidecomments
    \newcommand{\ksb}[1]{}
    \newcommand{\shane}[1]{}
    \newcommand{\jackson}[1]{}
    \newcommand{\melissa}[1]{}
    \newcommand{\dongwhi}[1]{}
    \newcommand{\koustuv}[1]{}
\else
    \newcommand{\shane}[1]{\textbf{\small\sffamily{\textcolor{shaneColor}{[#1 -- Shane]}}}}
    \newcommand{\jackson}[1]{\textbf{\small\sffamily{\textcolor{DarkBlue}{[#1 -- Jackson]}}}}
    \newcommand{\melissa}[1]{\textbf{\small\sffamily{\textcolor{dolive}{[#1 -- Melissa]}}}}
    \newcommand{\dongwhi}[1]{\textbf{\small\sffamily{\textcolor{dpink}{[#1 -- Dong Whi]}}}}
    \newcommand{\ksb}[1]{\textbf{\small\sffamily{\textcolor{marroon}{[#1 -- Karthik]}}}}
    \newcommand{\koustuv}[1]{\textbf{\small\sffamily{\textcolor{violet}{[#1 -- Koustuv]}}}}
  \fi

\newcommand{\kriya}{\texttt{KRIYA}}

\newcommand{\icheart}{\includegraphics[height=0.12in]{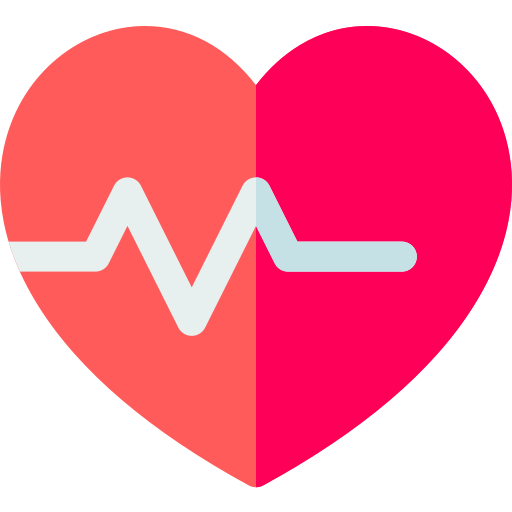}}
\newcommand{\icstep}{\includegraphics[height=0.12in]{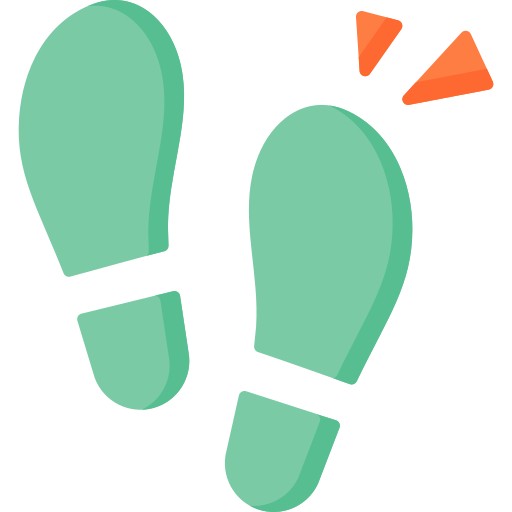}}
\newcommand{\icsleep}{\includegraphics[height=0.12in]{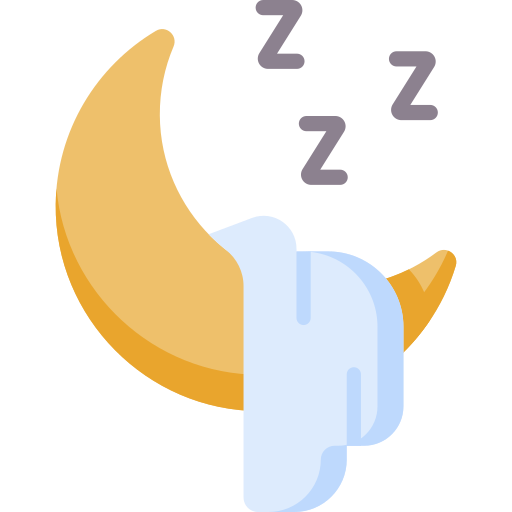}}
\newcommand{\icmind}{\includegraphics[height=0.12in]{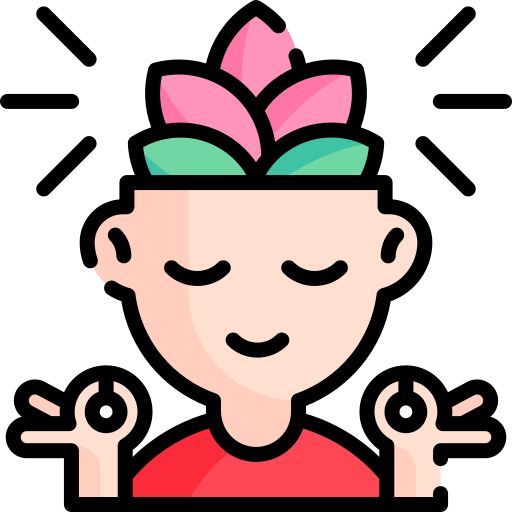}}
\newcommand{\icexercise}{\includegraphics[height=0.12in]{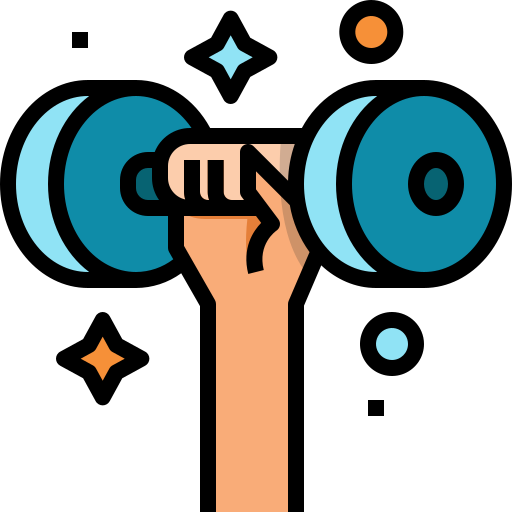}}
\newcommand{\icdiet}{\includegraphics[height=0.12in]{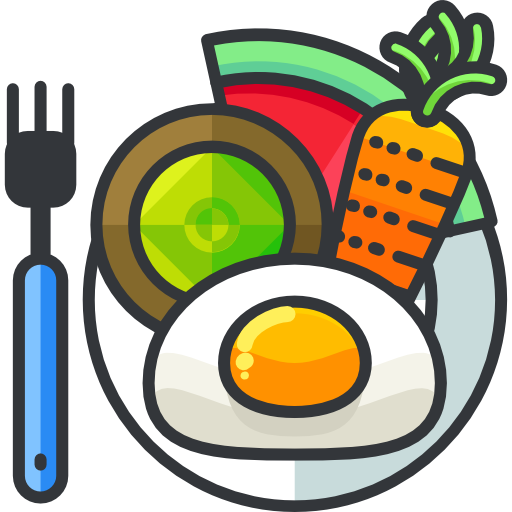}}

\renewcommand{\textrightarrow}{$\rightarrow$}

\newcommand{\n}[1]{$\mathtt{#1}$}

\colorlet{tableheadcolor}{gray!25} 
\colorlet{tablerowcolor}{gray!5} 
\definecolor{shaneColor}{HTML}{1F77B4}
\definecolor{neutralCol}{HTML}{dd1c77}
\definecolor{neutralGreen}{HTML}{31a354}
\definecolor{NewBlue}{HTML}{1879ba}
\definecolor{bleudefrance}{rgb}{0.19, 0.55, 0.91}  
\definecolor{AfTrColor}{HTML}{0868ac}  
\definecolor{BfTrColor}{HTML}{a8ddb5}  

\definecolor{AfCtColor}{HTML}{b10026}  
\definecolor{BfCtColor}{HTML}{fd8d3c}

\graphicspath{ {figures/} }

\newcommand{\para}[1]{\vspace{0.3em}\noindent\textbf{#1}~}

\AtBeginDocument{%
  \providecommand\BibTeX{{%
    \normalfont B\kern-0.5em{\scshape i\kern-0.25em b}\kern-0.8em\TeX}}}

\begin{document}


\title[Designing \kriya{}: An AI Companion for Wellbeing Self-Reflection]{Designing \kriya{}: An AI Companion for Wellbeing Self-Reflection}


\author{Shanshan Zhu}
\orcid{0009-0009-1191-7432}
\authornote{Both authors contributed equally.}
\affiliation{%
  \institution{University of Illinois Urbana-Champaign}
 \city{Urbana}
 \state{IL}
 \country{USA}}
 \email{szhu50@illinois.edu}

\author{Wenxuan Song}
\orcid{0009-0006-0718-6882}
\authornotemark[1]
\affiliation{%
  \institution{University of Illinois Urbana-Champaign}
 \city{Champaign}
 \state{IL}
 \country{USA}}
 \email{js129@illinois.edu}

\author{Jiayue Melissa Shi}
\orcid{0009-0007-0624-2421}
\affiliation{%
  \institution{University of Illinois Urbana-Champaign}
 \city{Urbana}
 \state{IL}
 \country{USA}}
 \email{mshi24@illinois.edu}

\author{Dong Whi Yoo}
\orcid{0000-0003-2738-1096}
\affiliation{%
 \institution{Indiana University Indianapolis}
 \city{Indianapolis}
 \state{IN}
 \country{USA}}
 \email{dy22@iu.edu}

\author{Karthik S. Bhat}
\orcid{0000-0003-0544-6303}
\affiliation{%
  \institution{Drexel University}
 \city{Philadelphia}
 \state{PA}
 \country{USA}}
 \email{ksbhat@drexel.edu}

\author{Koustuv Saha}
\orcid{0000-0002-8872-2934}
\affiliation{%
  \institution{University of Illinois Urbana-Champaign}
  \city{Urbana}
  \state{IL}
  \country{USA}}
\email{ksaha2@illinois.edu}

\renewcommand{\shortauthors}{}



\begin{abstract}
Most personal wellbeing apps present summative dashboards of health and physical activity metrics, yet many users struggle to translate this information into meaningful understanding. These apps commonly support engagement through goals, reminders, and structured targets, which can reinforce comparison, judgment, and performance anxiety. To explore a complementary approach that prioritizes self-reflection, we design \kriya{}, an AI wellbeing companion that supports co-interpretive engagement with personal wellbeing data. \kriya{} aims to collaborate with users to explore questions, explanations, and scenarios through features such as Comfort Zone, Detective Mode, and What-If Planning. We conducted semi-structured interviews with 18 college students interacting with a \kriya{} prototype using hypothetical data. Our findings show that through \kriya{} interaction, users framed engaging with wellbeing data as interpretation rather than performance, experienced reflection as supportive or pressuring depending on emotional framing, and developed trust through transparency. We discuss design implications for AI companions that support curiosity, self-compassion, and reflective sensemaking of personal health data.

\end{abstract}

\ccsdesc[300]{Human-centered computing~Empirical studies in collaborative and social computing}
\ccsdesc[300]{Applied computing~Psychology}

\keywords{wellbeing, digital wellbeing, personal health informatics, AI companion, compassionate design, co-interpretation}

\maketitle


\section{Introduction}\label{section:intro}

Contemporary personal health and wellbeing apps provide summaries and trends of steps, sleep, activity, and other metrics. 
While these tools offer precise, data-driven snapshots, many people struggle to stay motivated to engage with these dashboards or to translate their insights into meaningful action.
A majority of these designs rely on self-discipline-oriented strategies---such as goal setting, streaks, reminders, and self-tracking---or on intrusive nudges
that push users toward predetermined targets. 
These approaches often frame individuals as performers to be optimized, rather than as reflective interpreters of their own lived experience~\cite{baumer2014reviewing,sharon2017self,lupton2016diverse}.
Yet, personal wellbeing data are rarely straightforward; they fluctuate with context, emotion, environment, and daily variability such as weather and schedules.



Prior research in personal health informatics has examined how people collect, track, and make sense of wellbeing data for self-awareness and behavior change~\cite{sutton2016measuring}. 
However, most self-tracking systems continue to emphasize dashboards, progress indicators, goals, and reminders~\cite{li2010stage,epstein2015lived,fitzpatrick2013review}.
As a result, reflection is often framed as monitoring performance against predefined targets rather than as a process of interpretation. 
Research has shown that users frequently struggle to connect tracked metrics to lived experience and contextual factors such as stress, fatigue, or daily routines~\cite{loerakker2025framework,pantzar2017living}, and may disengage or abandon tools when tracking feels burdensome, judgmental, or misaligned with everyday realities~\cite{clawson2015no}. 
These limitations have motivated calls for more flexible, human-centered approaches to reflection---including narrative, annotation, and user-driven sensemaking---that move beyond purely metric-driven feedback~\cite{morris2018towards}.


A key challenge underlying these limitations is the distribution of interpretive and emotional labor.
Dashboards often offload the work of explanation---why metrics fluctuate, which contextual factors matter---onto users, even though such 
variability is common in everyday life~\cite{loerakker2025framework,pantzar2017living}. 
At the same time, performance-centered designs such as streaks, competition, and normative targets can intensify emotional labor, evoking 
guilt, anxiety, or a sense of failure when users fall short~\cite{peterson2022self,epstein2016beyond,lupton2016diverse,sharon2017self}. 
Together, these dynamics suggest the need for designs that help share the work of interpretation and emotional reframing, rather than simply increasing discipline or compliance.

This tension raises a critical design question: what if wellbeing technologies moved away from monitoring and motivating, and instead supported playful, co-interpretive engagement? 
Instead of judging or prescribing, what if systems could collaborate with individuals to interrogate patterns, explore possibilities, and uncover meaningful insights? 
Such a shift could reframe reflection from compliance to curiosity
inviting people to make sense of their data together \textit{with} a system, rather than being managed by it. 

Accordingly, we argue that co-interpretation can be an underexplored pathway for digital wellbeing that shifts reflection from compliance to collaborative sensemaking.
In the context of health tracking, co-interpretation could encourage an individual to remain 
curious about patterns while also considering contextual factors, such as weather, mood, schedule, and daily circumstances. 
Ideally, this reframing can shift the central question from ``Did I fail?'' to ``Why might this have happened?''---opening space for understanding rather than judgment. 
Conversational companions, in particular, may offer a promising way to scaffold such reflection by helping users articulate connections between data and context.
With the increasing excitement and development surrounding generative AI, recent research has explored generative AI-powered journaling (e.g., MindScape) to find that context-aware LLMs can act as reflective partners by embedding behavioral context into prompts and insights, reinforcing the promise of co-interpretive reflection~\cite{nepal2024mindscape}.

Despite this promise, little is known about how people experience conversational reflection around wellbeing data, or how AI-supported tools can scaffold it without creating new worries. 
To this end, we examine how an envisioned AI companion---designed based on compassionate and collaborative self-reflection---might reshape the experience of engaging with personal wellbeing data. 
Our work is guided by the following research questions (RQs):

\begin{enumerate}
    \item[\textbf{RQ1:}] How can an AI companion support more collaborative and emotionally attuned wellbeing self-reflection?
    \item[\textbf{RQ2:}] What concerns and desires do individuals express when engaging with an AI companion for wellbeing reflection?
\end{enumerate}

To explore our RQs, we designed a prototype~\cite{hutchinson2003technology}, called \kriya{}, an AI companion integrated into an envisioned health app dashboard.
\kriya{} embodies a design vision centered on compassionate, reflective, and emotionally attuned co-interpretation of personal wellbeing information. 
It scaffolds reflection through 
conversational interactions of features such as \textit{Morning Forecasts}, \textit{Evening Debriefs}, and \textit{What-If} scenarios. 
These interactions aim to make context visible and to support meaning-making instead of purely monitoring.

We scoped our study to young adults---specifically college students---a population deeply engaged with digital health and quantified-self technologies~\cite{knowles2025screenagers}. We conducted semi-structured interviews with 18 college students to examine how participants responded to hypothetical wellbeing scenarios grounded in everyday lived contexts.
Interviews included a guided prototype walkthrough in which participants interacted with \kriya{} using hypothetical data and reflected on how its conversational framing compared to more familiar health dashboards and tracking tools.
We applied thematic analysis to identify moments when the system supported curiosity-driven reflection, as well as points where participants raised concerns related to trust, interpretability, accuracy, and appropriate boundaries.



Our participants rated \kriya{} above accepted benchmarks on both the System Usability Scale (SUS; median=76.25) and the Intervention Appropriateness Measure (IAM; median=14).
Our qualitative analysis surfaced three central themes: (1) how \kriya{} supports a shift from passive tracking to active interpretation, (2) the role of compassionate reframing in shaping reflective engagement, and (3) factors influencing trust, understanding, and continued use.
Our findings highlight how conversational, co-interpretive designs can reshape users' relationships with their wellbeing data. Participants described moments where curiosity replaced self-judgment, and where contextual explanations helped normalize fluctuations that would otherwise feel like personal failure. 
They also articulated boundaries and concerns, revealing the delicate balance required when AI systems participate in reflective and emotionally sensitive practices. 
Together, these findings inform design implications for AI-mediated reflection tools that aim to cultivate curiosity and self-compassion while respecting users’ autonomy, interpretive agency, and emotional boundaries in personal health contexts.

\section{Related Work}


\subsection{Personal Health Informatics and Self-Reflection}


Personal health informatics research has investigated how people collect, track, and make sense of wellbeing data to support self-awareness and behavior change~\cite{rapp2016personal, sutton2016measuring,kersten2017personal}. 
Self-tracking systems typically emphasize visual summaries, progress indicators, goals, and reminders that encourage adherence to target behaviors and daily routines~\cite{li2010stage, epstein2015lived, fitzpatrick2013review}. 
These tools conceptualize reflection primarily as monitoring, checking whether one’s metrics meet predefined goals, rather than supporting deeper interpretive meaning-making. 
Prior work has explored various technologies to support self-tracking~\cite{arnera2024digital, kim2016timeaware, pammer2021reflection,karkar2016framework,feustel2018people,kim2017omnitrack,ayobi2018flexible,kirchner2021they}. 
Related work on digital self-control and screen use further illustrates the limitations of monitoring- and enforcement-oriented approaches to self-reflection. 
\citeauthor{roffarello2023achieving} conducted a systematic review and meta-analysis and found that digital self-control tools such as self-timers yield only small to medium effects in reducing time spent on distracting technologies~\cite{roffarello2023achieving}.
Similarly,~\citeauthor{zhang2022monitoring} found that dashboards and pop-up nudges can increase awareness of screen use but often fail to help users feel meaningfully in control~\cite{zhang2022monitoring}.

Consequently, researchers have critically examined how reflection is supported by the design of personal health informatics systems~\cite{ahmadpour2017information, kersten2017personal,khovanskaya2013everybody,epstein2020mapping, cecchinato2019designing}. 
Work on reflective informatics argues that many tools conflate reflection with compliance, offering limited support for interpretation, curiosity, or sensemaking beyond numerical feedback~\cite {baumer2014reviewing}. 
These systems often prioritize efficiency, correctness, and goal attainment~\cite{mccrickard2003model, horsky2012interface}. 
Empirical studies further show that users frequently struggle to connect tracked metrics to lived experience, contextual factors, or subjective states such as stress, fatigue, or motivation~\cite{loerakker2025framework, pantzar2017living}. 
Further, users may disengage, selectively attend to data, or abandon tools altogether~\cite{clawson2015no, bhat2020sociocultural, choe2014understanding}. 
Researchers have also critiqued how metric-driven designs can promote self-surveillance and moralized interpretations of behavior, framing deviations from targets as personal failure rather than reflections of contextual variability~\cite{sharon2017self}. 

Also, researchers have called for more flexible and human-centered approaches to self-reflection. Studies of informal and qualitative self-tracking highlight how narrative, annotation, and user-driven practices allow individuals to construct meaning in ways that better align with their goals and everyday realities~\cite{morris2018towards}. For example, \citeauthor{bhat2025my} demonstrates how lightweight, narrative-based reflection on everyday digital behaviors---such as estimating and reflecting on smartphone use---can support self-awareness and more intentional engagement~\cite{bhat2025my}. 
Together, this body of work 
motivates us to 
move beyond monitoring-oriented dashboards toward designing tools that support interpretation, context, and subjective sensemaking as central components of wellbeing reflection.
Our paper provides preliminary empirical insights into concerns and desires surrounding the conversational co-interpretation of personal health data.


\subsection{Emotional Attunement and Compassionate Design in Digital Wellbeing}

A growing body of work emphasizes emotional attunement and compassionate design as critical dimensions of digital wellbeing tools. Researchers have documented how performance-centered features---such as streaks, competition, and normative targets---can evoke guilt, anxiety, or perceived failure when users fall short of goals~\cite{peterson2022self}. 
Related critiques of self-tracking systems argued that metric-driven designs can lead to self-surveillance and moralized interpretations of behavior, undermining autonomy, motivation, and long-term engagement~\cite{epstein2016beyond, lupton2016diverse}.

In contrast, compassionate design approaches emphasize non-judgmental language, reflective prompts, emotional validation, and storytelling to foster psychological safety and self-compassion~\cite{van2023role}. 
\citeauthor{van2023role}'s systematic review
conceptualized compassionate technology as systems that explicitly acknowledge emotional struggle, normalize imperfection, and prioritize alleviating suffering over performance optimization~\cite{van2023role}. 
Likewise, \citeauthor{lusi2025designing} investigated how compassionate experiences are cultivated in mental health and wellbeing technologies, identifying design strategies---including storytelling, reflective rituals, and self-compassion practices---that support emotionally supportive engagement~\cite{lusi2025designing}.
Building on this foundation,~\citeauthor{ludden2024compassionate} situated compassion as a core value in the design of digital mental health technologies, articulating how value-based design practices can support emotionally attuned interactions rather than reductive behavioral metrics~\cite{ludden2024compassionate}.


Systems designed to support emotional reflection aim to shift users from self-surveillance toward gentler curiosity about internal states and lived experiences~\cite{baumer2014reviewing}. 
However, many existing approaches remain disconnected from an individual's personal data or provide only superficial mood tracking rather than integrated emotional interpretation.
We draw from this body of research by embedding emotionally attuned narrative language directly into the sensemaking of
personal wellbeing data---supporting reflection that acknowledges lived complexity and variability.
Our study extends compassionate design work by examining how users experience emotional resonance, safety, and meaning when interacting with an AI companion positioned explicitly as a reflective collaborator.

\subsection{Human-centered AI and Conversational Agents for Wellbeing}

Human-centered AI research emphasizes systems that prioritize user goals, agency, transparency, and meaningful interaction over automation-centric performance metrics~\cite{amershi2019guidelines,shen2022human,ehsan2023charting}. 
In wellbeing contexts, conversational agents have emerged as lightweight and approachable interfaces for reflection, emotional support, journaling, psychoeducation, and coaching~\cite{fitzpatrick2017delivering, inkster2018empathy, chang2024ai,sharma2024facilitating,das2025ai,yoo2024missed}. 
Compared to form-based or dashboard-driven tools, conversational modalities are often experienced as more natural and less intimidating, particularly when engaging with sensitive or emotionally charged topics~\cite{ho2018psychological}.
Prior work has shown that individuals may disclose sensitive mental health concerns to AI systems, especially in situations where human support is unavailable or stigmatized~\cite{shi2025mapping, croes2024digital}. 
These systems are frequently positioned as adjuncts or entry points to wellbeing support, offering advantages such as scalability, immediacy, and low-cost access~\cite{miner2016smartphone, chen2020creating,lai2023psyllm}.

At the same time, research has underscored the importance of grounding wellbeing technologies in users' lived experiences through participatory and co-design approaches~\cite{kilfoy2024umbrella, malloy2022co, mcgovern2025use}. 
Such approaches highlight how contextual, emotional, and situational factors shape engagement with conversational systems, beyond what can be captured through outcome-focused or performance-driven evaluations alone.

A majority of wellbeing chatbots function primarily as coaches or instructors, delivering reminders, techniques, or problem-solving guidance aimed at behavioral correction or symptom reduction~\cite{casu2024ai, shi2025mapping, yoo2025ai}. 
For example,~\citeauthor{fitzpatrick2017delivering} described Woebot as a system that guides users through structured, skills-based interactions inspired by cognitive behavioral therapy, prompting exercises such as thought reframing, mood check-ins, and goal setting in response to reported symptoms~\cite{fitzpatrick2017delivering}. 
Similarly,~\citeauthor{vaidyam2019chatbots} characterized many mental health conversational agents as providing psychoeducation, screening, and directive support rather than open-ended emotional exploration. 
These designs largely maintain a directive stance, framing reflection as a means toward compliance with prescribed targets or therapeutic routines~\cite{schloss2022towards}. 
While such approaches can be effective for delivering structured interventions or short-term symptom management, they also leave open questions about how conversational systems might support emotional nuance, contextual interpretation, and collaborative reflection around lived experience.

In parallel, research in personal informatics has begun exploring how AI can support self-care, reflection, and meaning-making around personal data~\cite{wang2025exploring,shin2025planfitting,fang2024physiollm}. Recent studies examine generative AI tools for self-care~\cite{capel2024studying}, 
long-term AI-driven dialogue to support self-reflection~\cite{li2025introspectus}, and AI-generated representations that reframe how individuals interpret personal data~\cite{park2025reimagining}. 
Together, this body of research signals a shift from tracking and visualization toward interpretation and reflection, raising questions about whether---and how---AI might meaningfully support this process.
This paper examines how a human-centered AI companion, \kriya{}, supports the co-interpretation of personal health data by engaging users in dialogue around emerging patterns rather than presenting conclusions in isolation.
\section{Study Design and Methods}\label{section:design}

\begin{figure*}[t]
    \centering
    \includegraphics[width=1.6\columnwidth]{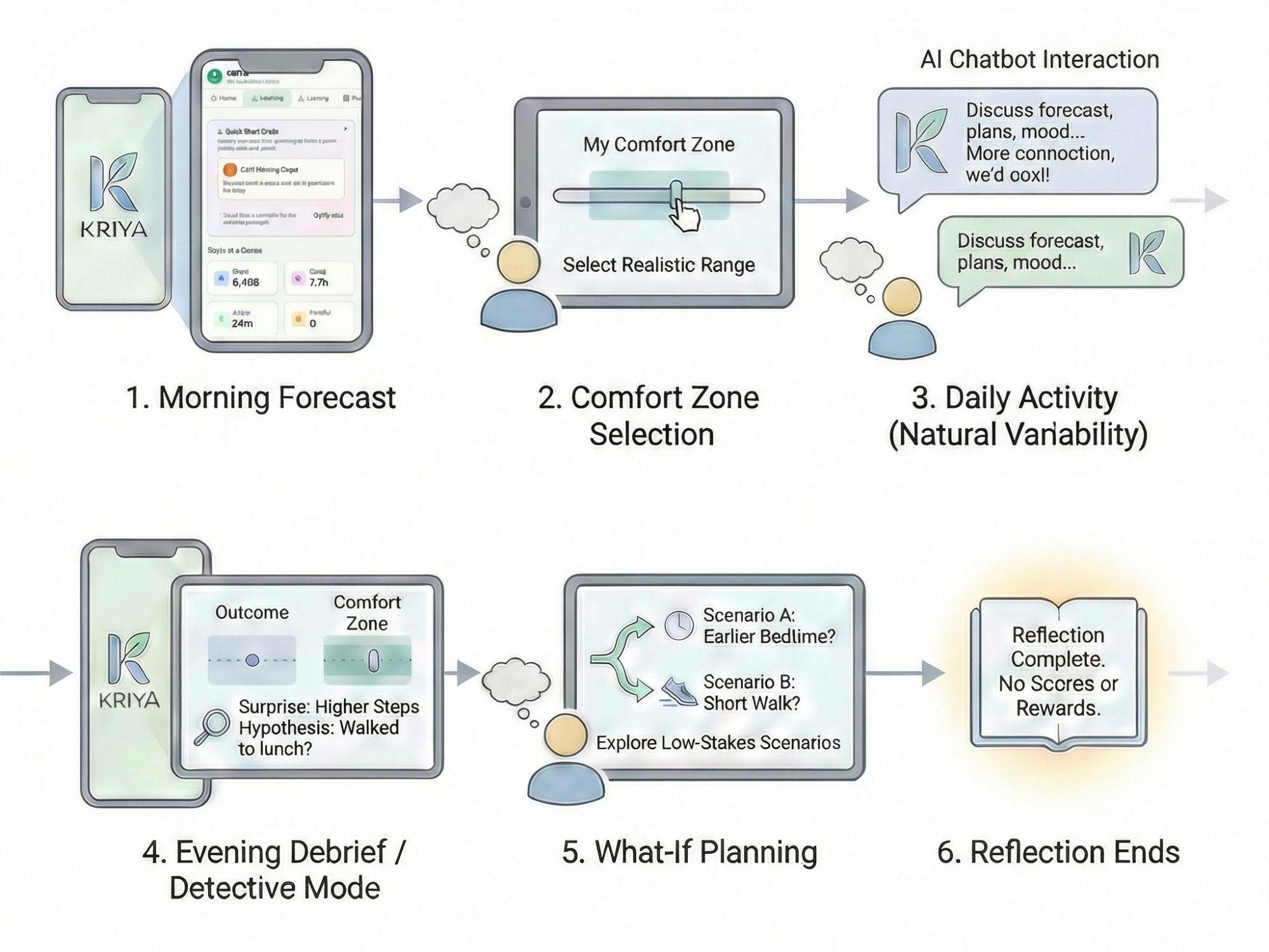}
    \caption{Interaction workflow of KRIYA (an AI companion prototype) during the interview study. The participants engaged with hypothetical data through a sequence of interactions, including 1) Morning Forecast (contextualized overview of anticipated patterns for the day), 2) Comfort Zone selection (participants set more realistic boundaries for the day based on schedule and comfort), 3) conversational reflection (dialogue-based sensemaking around recent data patterns), 4) Evening Debrief (retrospective reflection on the day's experiences), 5) What-If Planning (exploration of hypothetical changes and future scenarios), and 6) final reflections (participants' meta-level evaluation of the experience in the session).}
    \label{fig:study-workflow}
    \Description[figure]{Workflow diagram showing the sequence of \kriya{} interactions used in the interview walkthrough with hypothetical data. The flow progresses through six steps: Morning Forecast, Comfort Zone selection, conversational reflection in chat, Evening Debrief, What-If Planning, and final reflections. The diagram emphasizes a day-oriented reflection process moving from forecasting and setting a realistic range to interpreting outcomes and exploring future scenarios.}
\end{figure*}

Our study was reviewed and approved by the Institutional Review Board (IRB) at our university. 
We conducted an exploratory qualitative study
using \kriya{} as a prototype to understand how people experience wellbeing reflection when an AI system is integrated as a co-interpreter~\cite{hutchinson2003technology} (see~\autoref{fig:study-workflow} for an overview of study workflow). 
Through these interviews, we explored participants' experiences along: how they interpreted and explained patterns during self-reflection, how emotionally supportive the process felt, and how much trust they placed in the interactions.
To reduce privacy risk and isolate interaction framing from personal disclosure, participants interacted with hypothetical scenarios and dummy step and sleep data. 
This choice further enabled consistent comparison across participants.
Each session combined baseline questions about participants' existing practices and mental models with a structured walkthrough of the \kriya{} prototype. 
In this section, we elaborate on our methodology.

\subsection{Designing a Prototype AI Companion: \kriya{}}\label{subsec:kriya}

For our study, we designed a prototype~\cite{hutchinson2003technology}, called \kriya{}, of an interactive AI-powered wellbeing companion designed to support co-interpretation of personal metrics and everyday context.
In comparison to the majority of personal app designs, which consist of dashboards, targets, and progress tracking, we designed \kriya{} to scaffold explanation-oriented reflection.
Essentially, this design aims to help
users connect numbers to plausible contextual factors and by encouraging low-stakes exploration of ``why'' and ``what-if'' questions around their data. 

\begin{figure*}[t!]
 \begin{subfigure}[b]{\columnwidth}
    \centering
    \includegraphics[trim={0 10cm 0 0},clip,width=\columnwidth]{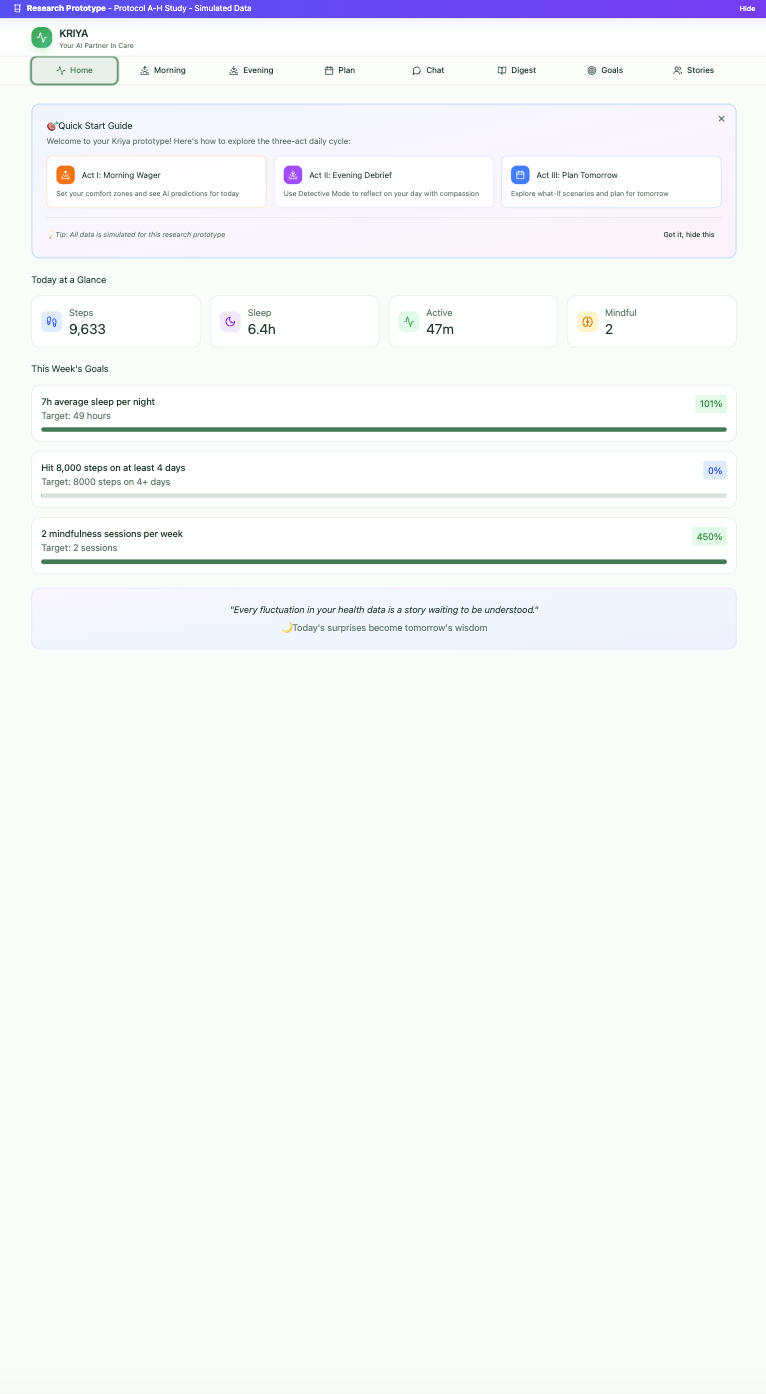}
    \caption{\kriya{} Dashboard}
    \end{subfigure}\hfill
     \begin{subfigure}[b]{\columnwidth}
    \centering
    \includegraphics[trim={0 10cm 0 0},clip,width=\columnwidth]{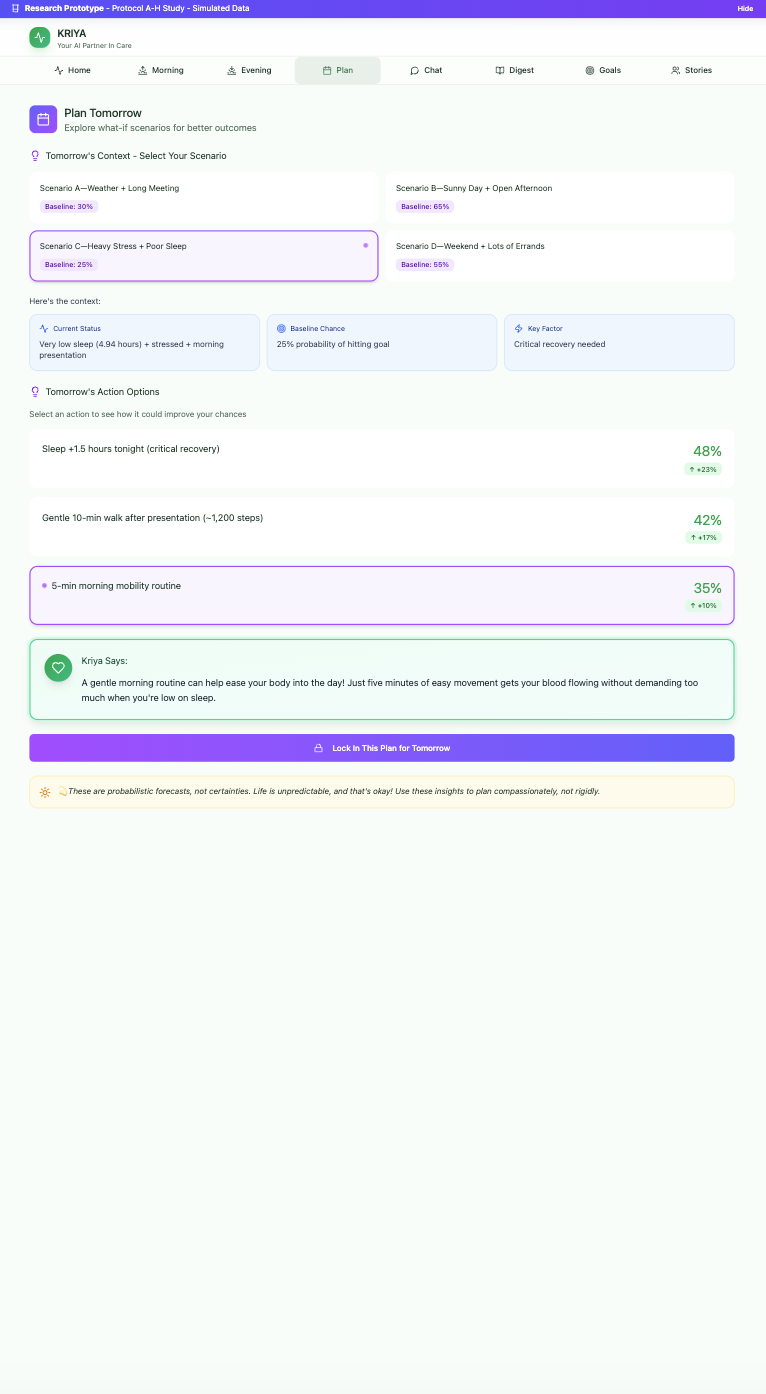}
    \caption{Plan}
    \end{subfigure}\hfill
    \caption{Screenshots of the AI companion prototype, \kriya{}, for the (a) dashboard and 
    (b) What-If Planning for speculative, low-stakes exploration of future scenarios.}
    \label{fig:kriya_screenshot}
    \Description[figure]{Two screenshots from the \kriya{} prototype. (a) The \kriya{} dashboard provides an overview of wellbeing information and entry points to reflective modules (e.g., forecasts, debriefs, and planning). (b) The What-If Planning screen supports low-stakes scenario exploration by showing a baseline likelihood of staying within a chosen Comfort Zone and offering small, feasible actions that can change the predicted likelihood, helping users compare options without framing them as strict goals.}
\end{figure*}

\autoref{fig:kriya_screenshot} and Appendix~\autoref{fig:kriya_morning} and \autoref{fig:kriya_evening} show some example screenshots of the \kriya{} prototype.
We designed \kriya{} with the aim of supporting reflection through four modules, as listed below: 

\para{A conversational chat space} where users ask questions about steps and sleep and receive context-linked responses. 

\para{A Morning Forecast module} that presents a baseline forecast informed by contextual signals (e.g., sleep, schedule, weather) and invites the user to select a Comfort Zone, defined as a realistic range rather than a single goal.


\para{An Evening Debrief module} with Detective Mode that compares outcomes to the chosen Comfort Zone, presents a Surprise Score indicating deviation from expectation, and offers possible factors to support explanation. 

\para{A What-If Planning module} that presents a baseline probability of staying within a future Comfort Zone and allows users to explore small, reversible alternatives (e.g., taking a short walk after class) to see how the probability might shift. 

Each participant was provided with a walkthrough of the \kriya{} tool. 
During the walkthrough, participants interpreted hypothetical scenarios based on the provided context and data, and then reflected on how the same interactions would apply to their own lives (\autoref{fig:study-workflow}). 

\subsection{Participants and Recruitment}\label{subsec:participants}

For our study, we recruited 18 college students through social media recruitment, particularly from the subreddit of our university. 
We focused on young adults---specifically, college students---a demographic deeply engaged with and invested in digital health and quantified-self technologies, which makes them especially suitable for studying reflective sensemaking around wellbeing data~\cite{knowles2025screenagers}.
We posted our recruitment flyer with an interest form that included a demographic survey questionnaire (age, gender), student status, and health app usage in their daily lives. Interested participants were asked to provide their institutional email IDs, enabling us to verify identities and filter out bots, duplicates, and fake responses from our interested participant pool.
Our inclusion criteria were that participants were 18 years or older, living in the US, and comfortable with discussing the use of personal health apps.


We screened the responses to identify a shortlist of a diversity of participants spanning a range of both frequent and non-users of health-tracking apps. 
In particular, we sought both frequent users and non-users because non-use can reveal when existing tools feel burdensome, discouraging, or low-value. 
Participants received a \$15 USD gift card for a 60-minute interview session. 
In total, we recruited 18 participants, who were interviewed between late-November 2025 and early-January 2026. 
\autoref{tab:demographics} summarizes the demographics and health app use experience of the participants.


\begin{table*}[t!]
\centering
\sffamily
\footnotesize
\caption{Summary of the 18 participants enrolled in the study, including age range, gender, highest level of education, and experience with health apps (e.g., Apple Health, Google Fit, Fitbit) for Step counts/activity \icstep, sleep \icsleep{}, heart rate \icheart{}, mindfulness/meditation \icmind{}, exercise \icexercise{}, and diet/nutrition \icdiet{}.}
\label{tab:demographics}
\begin{tabular}{lllllll}
 &  &  &  & \multicolumn{3}{c}{\textbf{Health App Usage}}\\
 \cmidrule(lr){5-7}
\textbf{ID} & \textbf{Age (Ys)} & \textbf{Gender} & \textbf{Highest Education} & \textbf{How Long?} & \textbf{How Frequent?} &  \textbf{Types}\\
\toprule
P1  & 18--24 & Woman & Some college, no degree & 6--12 months & Multiple times per day  &
    \icstep{} \icsleep{} \icheart{} \icexercise{} \icmind{}\\
\rowcollight P2  & 18--24 & Man & Some college, no degree & 6--12 months & A few times per week  & \icstep{} \icheart{}\\
P3  & 18--24 & Woman & Some college, no degree & <6 months & A few times per week	&  \icstep{} \icsleep{} \icheart{}\\
\rowcollight P4  & 18--25 & Man & Some college, no degree & Never used & Never &  \icexercise{}\\
P5  & 18--26 & Man & Bachelor’s & 2--5 years & A few times per week &  \icstep{} \icdiet{}\\
\rowcollight P6  & 18--27 & Woman & High school & 2--5 years & A few times per week	&  \icstep{} \icexercise{} \icdiet{}\\
P7  & 18--28 & Man & High school & 1--2 years & Multiple times per day	&  \icstep{} \icsleep{} \icheart{}\\
\rowcollight P8  & 18--29 & Man & High school & <6 months & Once per day &	 \icstep{}\\
P9  & 18--30 & Man & High school & Never used & Rarely	& \icexercise{}\\
\rowcollight P10 & 25--34 & Woman & Master's & 6--12 months & Rarely	&  \icstep{} \icsleep{} \icheart{}\\
P11 & 18--32 & Man & High school & Never used & Rarely	& \icsleep{}\\
\rowcollight P12 & 18--33 & Woman & Some college, no degree & 6--12 months & Rarely	&  \icstep{} \icsleep{}\\
P13 & 25--34 & Prefer not to say & Bachelor's & 1--2 years & Once per day	&  \icstep{} \icsleep{}\\
\rowcollight P14 & 25--34 & Man & Master's & 1--2 years & Multiple times per day	&  \icstep{} \icexercise{} \icdiet{}\\
P15 & 25--34 & Man & Bachelor’s & <6 months & A few times per week	&  \icstep{} \icsleep{} \icheart{}\\
\rowcollight P16 & 18-24 & Man & High school  & 2--5 years & Multiple times per day &  \icstep{} \icheart{} \icexercise{}\\
P17 & 25--34 & Man & Bachelor’s & Never used & Once per day	&  \icstep{} \icexercise{}\\
\rowcollight P18 & 18-24 & Man & Some college, no degree & 2--5 years & A few times per week &  \icstep{} \icsleep{} \icheart{} \icexercise{}\\
\end{tabular}
\Description[table]{This table summarizes demographic characteristics of the study participants, including age range, self-reported gender identity, highest educational attainment, and duration of prior experience with wellbeing or mental health technologies.}
\end{table*}

\subsection{Study Procedure}\label{subsec:procedure}
Before the interview, participants completed an entry survey, consisting of validated trait-based questionnaires: 1) personality traits through BFI-10~\cite{soto2017next}, 2) self-regulation skills with respect to digital use adapted
from the Short Self-Regulation Survey Questionnaire (SSRQ)~\cite{carey2004psychometric},
3) self-control traits through the Brief Self-Control scale (BSCS)~\cite{tangney2018high}.
\autoref{table:individual_differences} summarizes the descriptive statistics of personality traits, self-regulation, and self-control of the participants.
We find that our participant pool shows a substantial variability in both self-regulation (SSRQ mean=46.04, stdev.=6.19) and self-control (BSCS mean=3.43, stdev.=0.65).

\begin{table}[t]
\sffamily
\centering
\footnotesize
\caption{Descriptive statistics of self-reported individual differences of the \n{N}=18 participants enrolled in the study. The distributions include the range, mean (\n{\mu}), and standard deviation (\n{\sigma}).}

\resizebox{\columnwidth}{!}{
\begin{tabular}{lllr}
    \textbf{Individual Difference} & \textbf{Distribution} \\
    \cmidrule(lr){1-1} \cmidrule(lr){2-4}
    \rowcollight \multicolumn{4}{l}{\textit{Personality Trait (BFI scale)}} \\
    ~Extraversion & [2, 4.5], \n{\mu}=2.94, \n{\sigma}=0.78 &
    \includegraphics[width=0.45in]{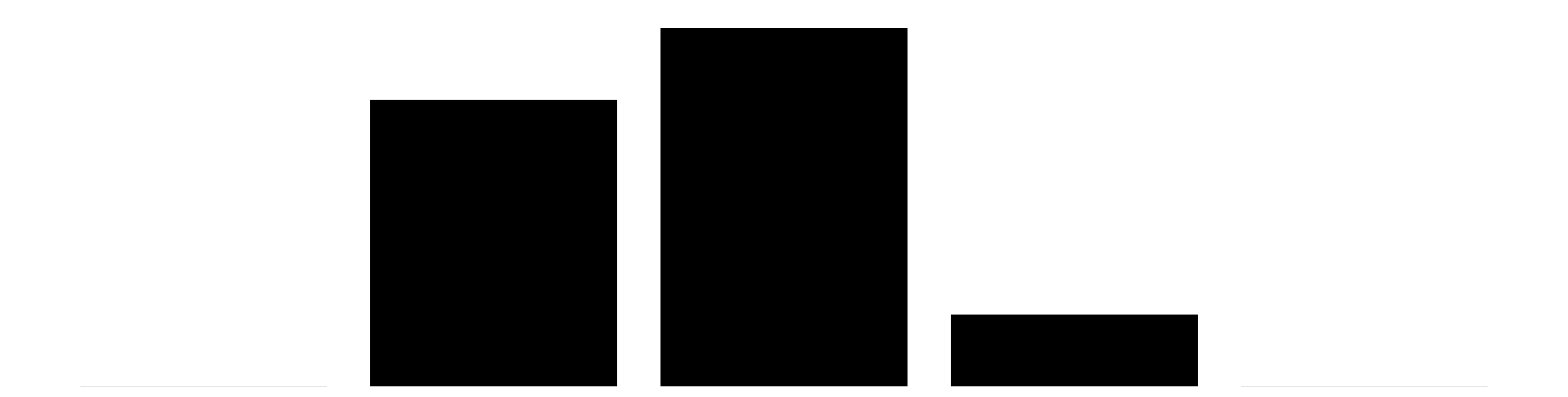} \\
    ~Agreeableness & [2, 5], \n{\mu}=3.5, \n{\sigma}=0.90 &
    \includegraphics[width=0.45in]{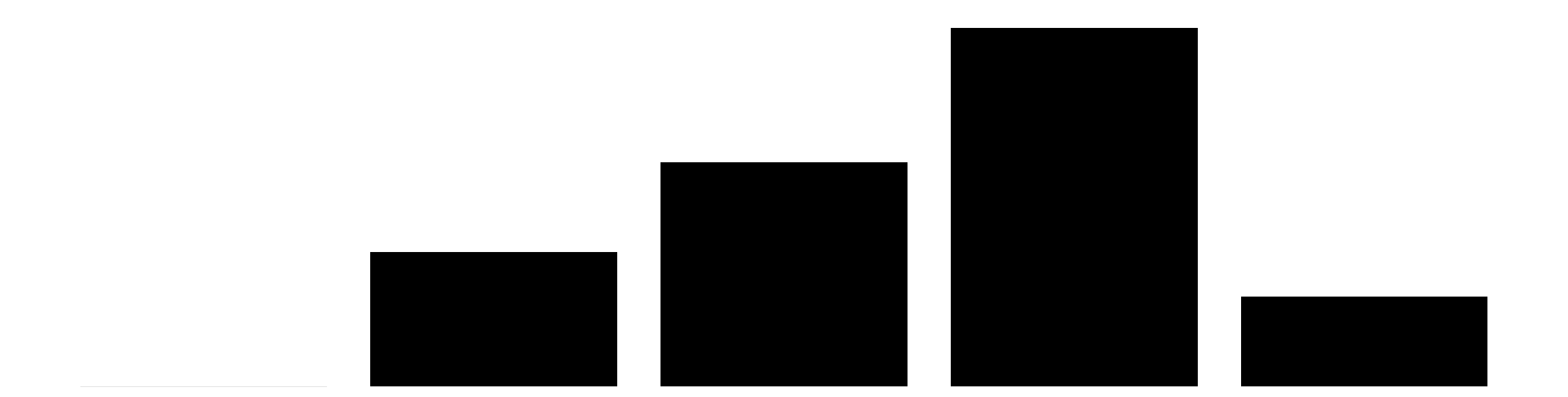} \\
    ~Conscientiousness & [2, 4.5], \n{\mu}=3.25, \n{\sigma}=0.71 &
    \includegraphics[width=0.45in]{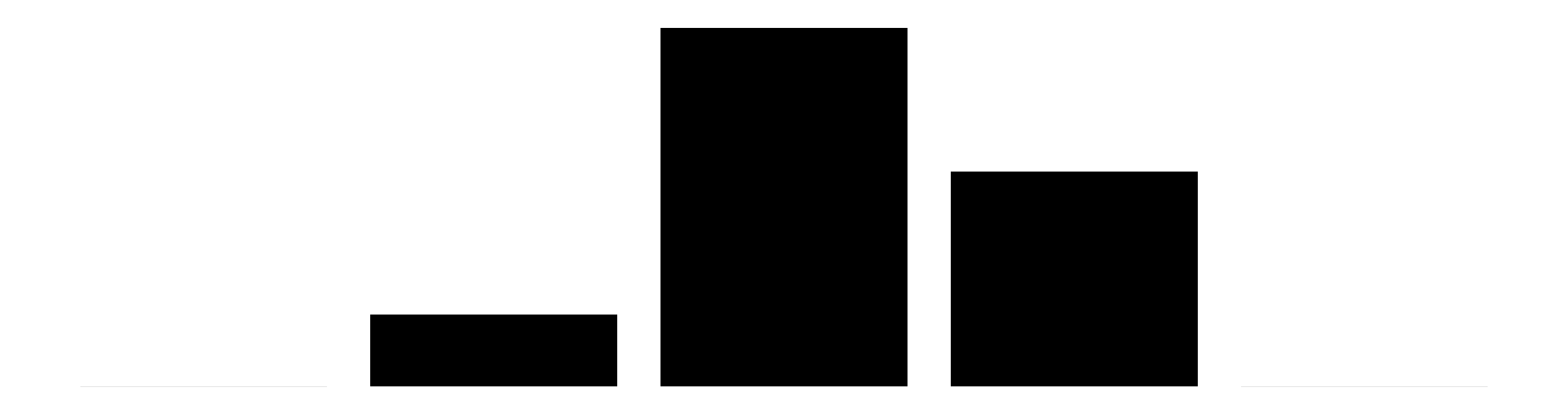}\\
    ~Neuroticism & [2, 4], \n{\mu}=3.11, \n{\sigma}=0.72 &
    \includegraphics[width=0.45in]{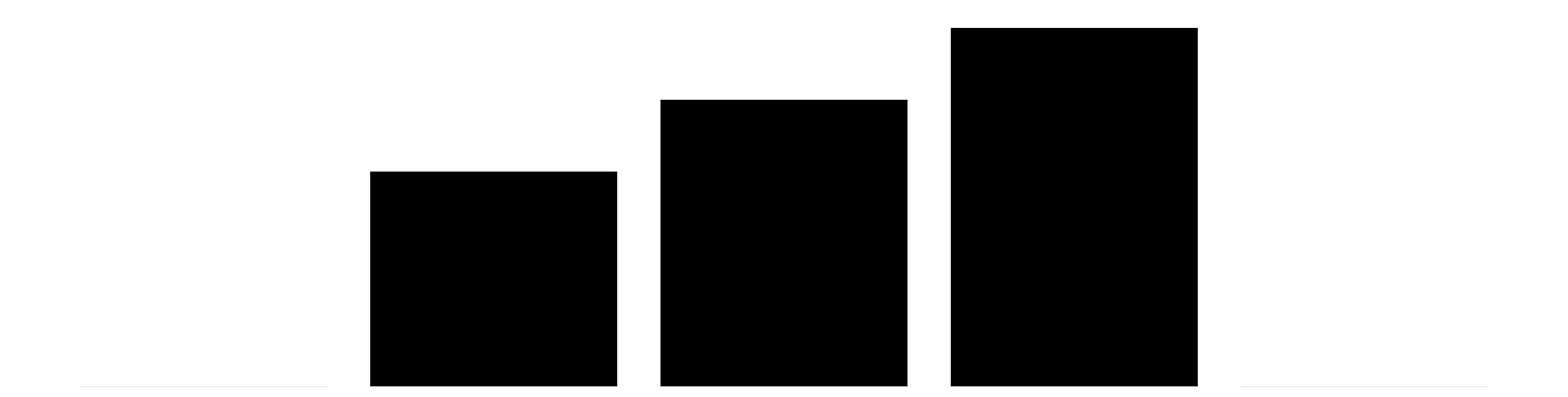} \\
   ~Openness & [1, 5], \n{\mu}=3.17, \n{\sigma}=1.13 &
    \includegraphics[width=0.45in]{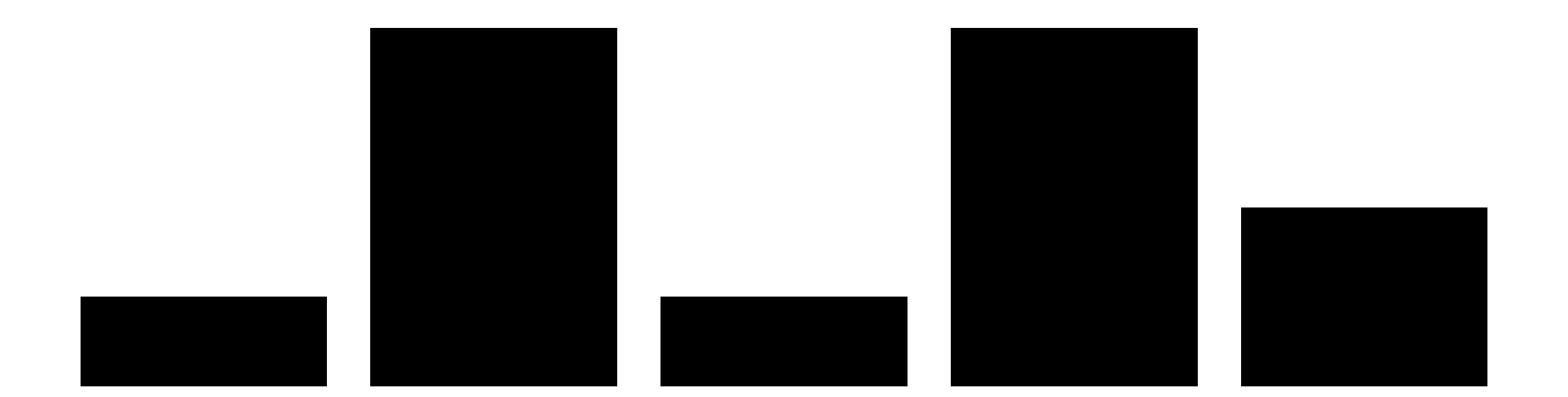} \\

    \rowcollight \multicolumn{4}{l}{\textit{Self-Regulation  and Self-Control}} \\
    ~Short Self-Regulation Questionnaire (SSRQ) & [27, 55], \n{\mu}=46.04, \n{\sigma}=6.19 &
    \includegraphics[width=0.45in]{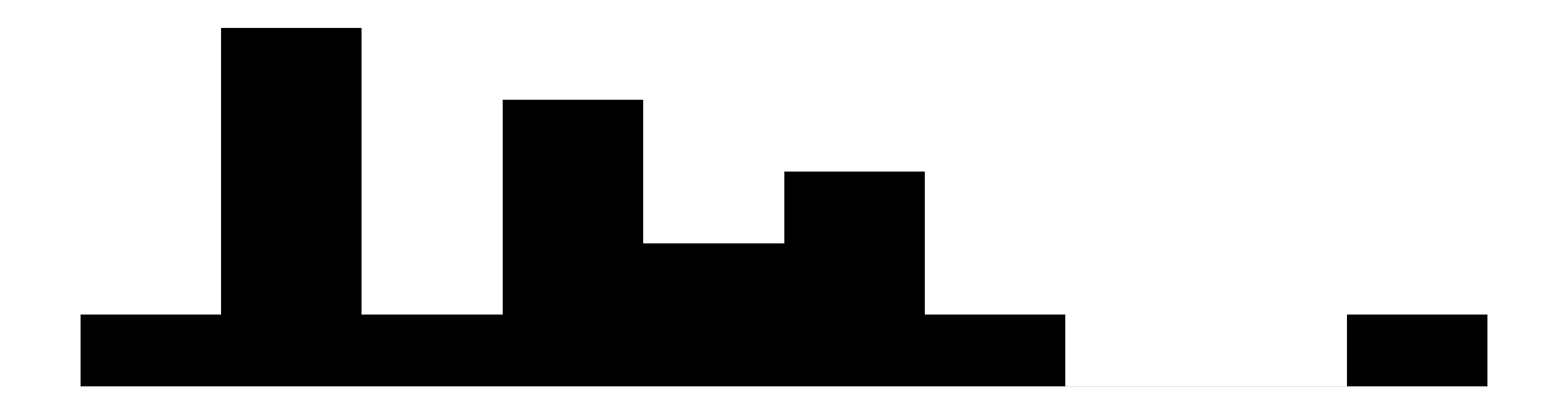} \\
    ~Basic Self-Control Scale (BSCS) & [1.7, 4.5], \n{\mu}=3.43, \n{\sigma}=0.65 &
    \includegraphics[width=0.45in]{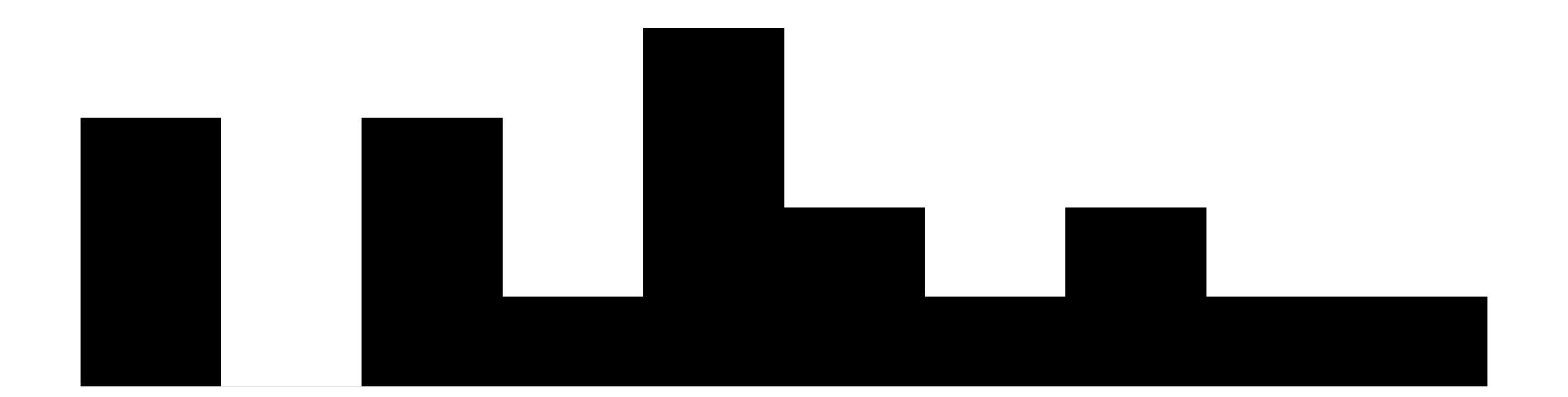} \\
\end{tabular}}
\label{table:individual_differences}
\Description[table]{This table summarizes the descriptive statistics of self-reported individual differences from the 18 participants in the study. These are organized into two categories: 1) personality traits (assessed using the BFI scale), and 2) self-regulation and self-control. For each measure, the table displays the observed range, mean (μ), standard deviation (σ), and a histogram illustrating the distribution across participants. Personality trait scores are calculated as item-level means on a 1 to 5 scale, while the self-regulation and self-control measures are based on summed or averaged scores across multiple items, resulting in broader value ranges determined by their respective instruments.}
\end{table}

Each participant was invited to join a 60-minute video call over Zoom. 
With consent, these interviews were recorded, automatically transcribed by Zoom, and manually verified for accuracy immediately after completion. 
Each interview session consisted of three dedicated segments that moved from current practices to \kriya{} interaction and reactions.

\subsubsection{Segment A: Baseline Usage and Mental Models.}\label{subsec:segmentA}
We asked participants how they use health data in daily life. 
Participants described how often they open health apps, and their motivations of using these apps (if at all).
If participants responded that they rarely or never check health apps, we asked about what factors and concerns shaped that choice. 
Further, we asked participants to estimate their typical step count and sleep duration without checking their smartphone or wearable app.
We also probed participants whether they encountered moments of surprise when viewing their activity or sleep data on these apps.
These questions captured a baseline sense of how participants think about their own patterns before they see any tracked values.

\subsubsection{Segment B: Checking Current Health App}\label{subsec:segmentB}
Participants who reported frequent app use opened their primary tool, such as Apple Health (on Apple devices) or Google Fit (on Android devices). 
Participants were asked to review their recent step and sleep data and compare it with their earlier estimates. They were then asked to describe whether the data aligned with their expectations or elicited surprise, and whether they could explain any discrepancies between their estimates and actual activity or sleep (e.g., due to schedule changes, stress, or weather).
We further asked participants to share their thoughts on the design and usability of these health app dashboards. 

\subsubsection{Segment C: \kriya{} Walkthrough and Post-Interaction Reflection}\label{subsec:segmentC}
In the next segment, the interview facilitator introduced \kriya{} as an experimental AI companion that supports co-interpretation of personal health data. 
Participants completed a guided walkthrough of \kriya{} using hypothetical scenarios and simulated data.
In particular, we used created four hypothetical scenarios in the \kriya{} platform---1) \textit{Scenario A: Rainy Weather + Long Meeting}, 2) \textit{Scenario B: Sunny Day + Open Afternoon}, 3) \textit{Scenario C: Heavy Stress + Poor Sleep}, and 4) \textit{Scenario D: Weekend + Lots of Errands}. 
Participants were asked to choose a scenario that they related most closely to for walking through the prototype.
They interacted with the conversational chat for 2--3 turns by choosing or typing questions about steps, sleep, and day-to-day variation. 
We asked how the AI's conversational tone felt, whether responses were understandable, and whether the chat supported reflection.

Participants explored one hypothetical Morning Forecast scenario and set a Comfort Zone for steps (and for sleep when available). 
We asked whether the system's reasoning felt clear and whether a range felt more natural than a single target. 
Participants then moved to the Evening Debrief scenario and used Detective Mode that matched their Morning Forecast selection. They interpreted the Surprise Score, evaluated proposed contributing factors, and described what felt realistic, too general, or missing. Participants also explored one What-If Planning scenario. They compared the baseline probability of staying within the next-day Comfort Zone with suggested actions, selected one action that felt doable, and described how the probabilistic framing shaped their planning.

After the walkthrough, participants reflected on their experiences with the interactions,
whether they would trust \kriya{} with real data, what privacy boundaries they preferred, what types of errors would be acceptable, and what changes would make the tool fit their routines.

At the end of the study, the participants also completed an exit survey assessing the perceived usability (SUS)~\cite{bangor2008empirical} and intervention appropriateness (IAM~\cite{weiner2017psychometric}) of \kriya{}.

\subsection{Data Analysis}\label{section:analysis}

After each interview, we manually verified the automatically generated transcripts to ensure accuracy and anonymized them by removing identifying details. 
We analyzed these transcripts using reflexive thematic analysis~\cite{braun2019reflecting}. The co-first authors led the initial round of open coding on the raw transcripts. 
The full author team met regularly to review selected excerpts, discuss differences in interpretation, and refine code definitions. 
This shared workflow helped us keep the coding approach aligned across participants and helped us agree on what kinds of excerpts supported a given claim. 
We initially obtained 425 codes, which were grouped into six lower-level subthemes and three higher-level themes that aligned with our research questions.
We kept a short lens sheet to anchor coding decisions to the study aims. 
While reviewing and coding the transcripts, we additionally wrote brief case memos for each participant that summarized their baseline practices with health apps, notable moments during the \kriya{} walkthrough, and tensions that shaped their reactions. 
As we iteratively revisited the corpus, we merged closely related codes, separated codes that captured different ideas, and consolidated the remaining codes into higher-level themes.

\subsection{Privacy, Ethics, and Positionality}\label{subsec:ethics}
Our study was reviewed and approved by the Institutional Review Board (IRB) at our university. 
We treated wellbeing reflection as sensitive data. We stored recordings, transcripts, and analysis files in access-controlled storage, and removed personally identifying information from transcripts before analysis. 
We also limited privacy risk through our study design: participants interacted with a hypothetical persona and dummy step and sleep data during the walkthrough. 
This approach reduced the need for participants to share personal health records while still allowing us to examine how they responded to conversational co-interpretation.

Our research team comprises members holding diverse gender, racial, and cultural backgrounds, including researchers who are people of color and immigrants, and spans interdisciplinary expertise across HCI, CSCW, human-centered design, and AI ethics. 
As a team, we have prior research experience in digital wellbeing and technology use. 
While we have taken care to faithfully capture and synthesize participants' perspectives, we acknowledge that our interpretations are situated and shaped by our disciplinary training, professional backgrounds, and personal experiences.
\section{Results}
Across interviews, participants frequently drew comparisons between \kriya{} and more familiar personal health dashboards, such as those from Google Fit, Apple Health, and Fitbit, using these contrasts to explain how \kriya{} shaped their understanding of data, emotional responses, and trust during reflection. 
In particular, participants described self-reflection as something that unfolded through dialogue, explanation, and the integration of everyday context.
In this section, we describe how participants made sense of \kriya{} through interaction. 
Throughout, we draw contrasts that participants used to ground their accounts, including \kriya{} versus familiar dashboards, steps versus sleep, and prior users versus non-users of wellbeing apps.
We first report on perceived usability and appropriateness (\autoref{sec:perceived_usability}), and then present three themes from our qualitative analysis: shifting from tracking to interpretation (\autoref{results:tracking_to_interpretation}), influence of compassionate framing for reflection (\autoref{results:emotional_framing}), and factors shaping trust, understanding, and continued use (\autoref{results:trust_understanding_use}).

\subsection{Perceived Usability of \kriya{}}\label{sec:perceived_usability}

\begin{figure}[t!]
\centering
\begin{subfigure}[b]{0.49\columnwidth}
    \centering
    \includegraphics[width=\columnwidth]{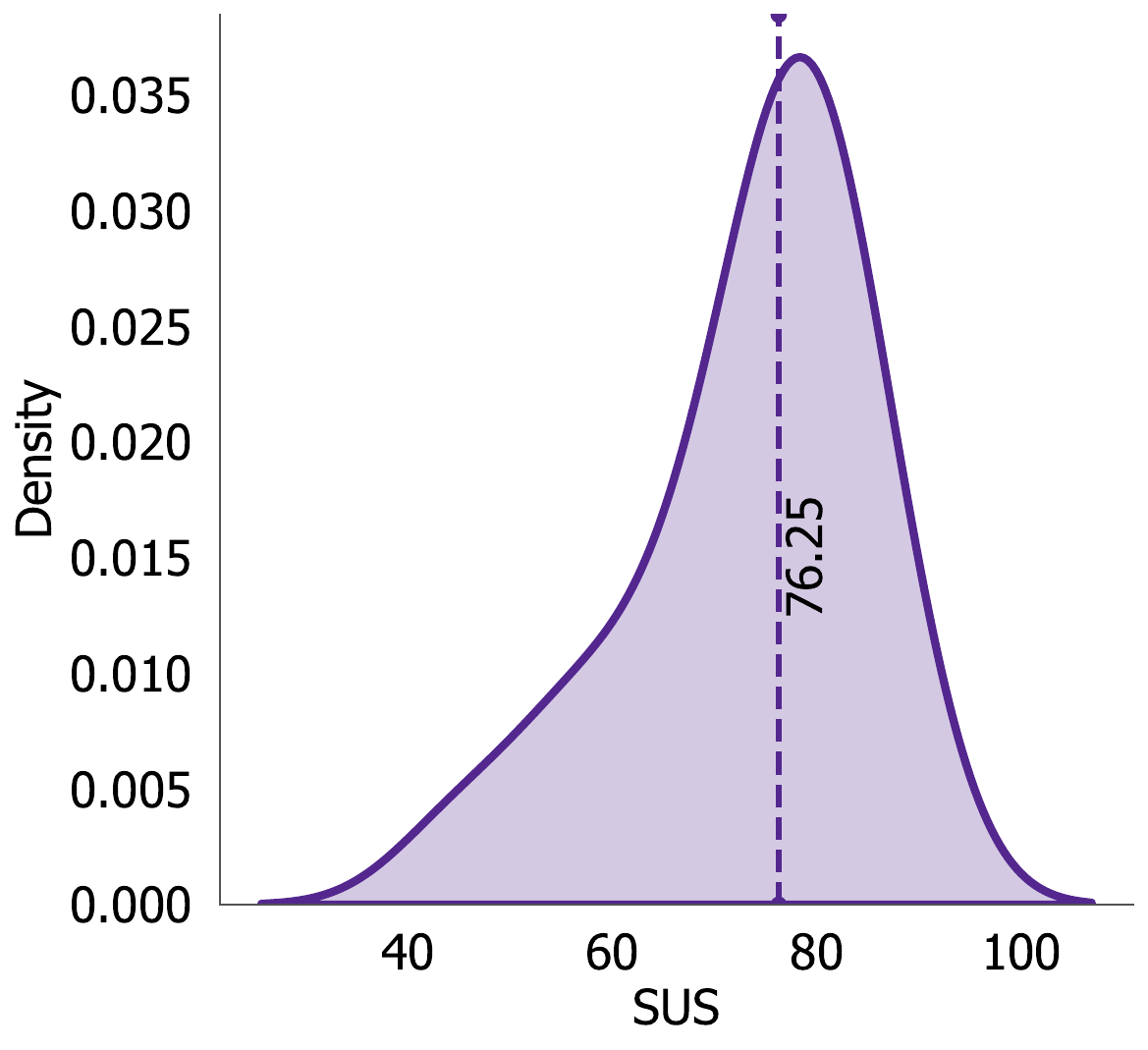}
    \caption{SUS}
    \label{fig:sus}
    \end{subfigure}
\begin{subfigure}[b]{0.49\columnwidth}
    \centering
    \includegraphics[width=\columnwidth]{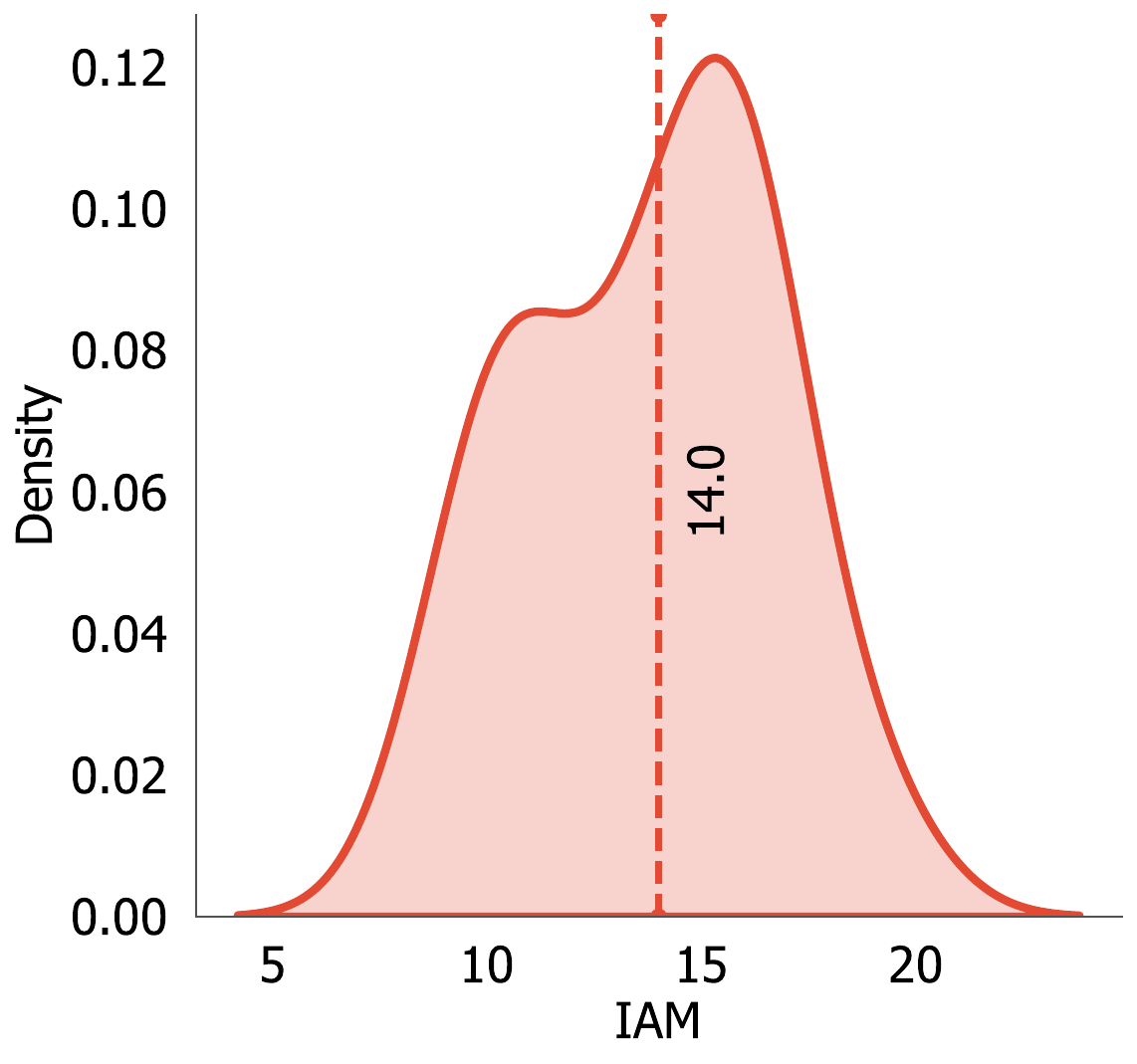}
    \caption{IAM}
    \label{fig:iam}
    \end{subfigure}\hfill
    \caption{Distribution plots based on the exit surveys (the dotted lines represent the median values).}
    \label{fig:dist_plots}
    \Description[figure]{This figure shows the distribution of self-reported scores from exit surveys completed by participants. Plots (a) and (b) show kernel density estimates for System Usability Scale (SUS) and Information Assessment Method (IAM) scores reported at the end of the study, with vertical dotted lines indicating the median values.  The corresponding dotted lines show the median of each distribution.}
\end{figure}



To begin with, we look at participants' responses to the exit surveys on usability and appropriateness of \kriya{}.

Participants completed the System Usability Scale (SUS; 0--100)~\cite{bangor2008empirical}. The median SUS score was 76.25 (stdev.=11.18), which falls above commonly cited thresholds for acceptable usability ($\sim$68~\cite{lewis2018item}).
This can be interpreted as the participants were 
largely able to navigate and engage with \kriya{} without substantial friction.

In addition, participants completed the Intervention Appropriateness Measure (IAM; 4--20)~\cite{weiner2017psychometric}, where they self-reported a median IAM score of 14.0 (stdev.=2.78), indicating that \kriya{} was perceived as a moderately appropriate intervention for wellbeing reflection. 
Consistent with prior work using the IAM~\cite{weiner2017psychometric}, this level of appropriateness suggests that participants viewed the system as a reasonable and potentially valuable fit for reflective practice, while also leaving room for improvement in how well it aligned with individual contexts or expectations. 

For instance, participants described the usefulness of \kriya{} in how it supported explanation-oriented reflection. 
Through conversational prompts and explanatory breakdowns, the system helped reflection feel less like monitoring performance and more like understanding everyday patterns. 
P12 explained the ``What If'' Planning module made change feel more manageable, noting that ``y\textit{ou don’t have to really make a big change in your lifestyle,}'' while the system's way of ``\textit{offering the options}'' highlighted how ``\textit{little things you can do}'' could still contribute meaningfully, making progress toward goals feel ``\textit{a little less daunting}''. 
P16 described the evening debrief's surprise breakdown as especially helpful, explaining that:

\begin{quote}
\small
``The five hours straight desk work and tagged stress in the afternoon [..] It seems to be helpful, and I think it was a good idea.'' (P16)
\end{quote}

It is also important to interpret these scores in light of the system's maturity and the study's sample size. 
\kriya{} was only used as a prototype during the interview session rather than a 
fully polished or deployed system. 
As such, the observed SUS and IAM scores not only point to the promise of this design approach for AI-supported wellbeing reflection but also highlight opportunities for further investigation and refinement.
That said, participants also voiced concerns with technical issues and glitches in \kriya{}. In particular, they noted that the system's explanations sometimes felt overly general and constrained by fixed assumptions, limiting its ability to reflect the shift circustances of everyday life. As P13 explained, ``The reasons might vary on a daily basis and these are more generic ones\dots I would like an option to add on so that it can make more informed decisions,'' while also noting that they understand the system was currently limited by predefined structures, adding that:

\begin{quote}
\small
``I understand that now it has a predefined list of things it’s working with, but I would like it to be much more engaging and conversational, like I can ask it anything and add on to the questions\dots not just based on the already fed prompts or things it has.'' (P13)

\end{quote}





\subsection{From Tracking to Interpretation: Data as an Invitation for Explanation}\label{results:tracking_to_interpretation}


Across the interviews, we observed an interesting shift in how participants perceived and interpreted wellbeing data through \kriya{}. 
Rather than approaching personal metrics as signals to be monitored, judged, or optimized, they engaged with them as prompts for explanation and sensemaking. 
We noted how the interaction with \kriya{} encouraged them to ask why particular patterns appeared, to reflect on how daily context shaped those patterns, and to treat data as something that required interpretation rather than evaluation. 
In this way, participants described reflection as an active process, unfolding through interaction. 
We elaborate this theme through two sub-themes: 1) \textit{Discrepancies lead to curiosity and inquiry}, and 2) \textit{Co-Interpretation explains discrepancies and supports self-reflection}, as described below:


\subsubsection{Discrepancies lead to curiosity and inquiry}
Participants described engaging with \kriya{} as a way to interpret wellbeing data rather than to judge performance. 
Multiple participants shared how reflection was often triggered when they noticed a mismatch between how they felt and what the data showed. 
They described moments of surprise when step or sleep values did not align with their estimates about the recent days. 
Rather than triggering self-criticism, these discrepancies commonly sparked curiosity. 


For example, P13 described noticing that their step count was lower than expected and beginning to reason through possible explanations: ``\textit{I think I walk more than what I see in this, [..] but when I see it, it's pretty much lower than what I was expecting},'' which led them to reflect on whether it was ``\textit{probably a holiday}'' where they stayed mostly at home, resulting in ``\textit{not much steps}''.

Several participants described this experience in relation to step counts. For example, P6, P8, and P10 noted that their perceived activity often felt higher than what their current data showed. P10 explained:

\begin{quote}
\small
``Sometimes I feel I walked a lot, but when I check, it's often fewer steps than what I would imagine. Like I would probably think I walked like 8000 steps, but when I open it, it's more likely just 6000 or something.'' (P10)
\end{quote} 

However, some participants also emphasized that surprise alone did not reliably lead to insight. 
When discrepancies were surfaced without explanation or interpretive guidance, they could feel confusing rather than helpful. As P1 noted in a later interaction:

\begin{quote}
\small
``It kind of feels a little bit disconnected from what I was wondering, like how can I meet my steps.'' (P1)
\end{quote}
This points to a boundary condition: when \kriya{} highlights a discrepancy without a clear explanation or next step, the surprise can feel disconnected from participants' goals and can weaken reflection.

\subsubsection{Co-interpretation explains discrepancies and supports self-reflection}
Building on the above need for explanation,
participants described reflection as something that unfolded through dialogue with \kriya{}, rather than through viewing numbers on their own. 
They appreciated that \kriya{} showed possible explanations and invited them to respond or elaborate. 
P2 explicitly contrasted this experience with conventional health apps, mentioning:

\begin{quote}
\small
``I think the reasoning step is very interesting to me, because it actually tells me why it underperformed instead of just showing me the number.'' (P2)
\end{quote}

P2 further elaborated that while the data patterns themselves were relatively straightforward, the accompanying explanation added value: ``\textit{The pattern itself was [..] very similar to just the data itself. So that was less interesting to me, but the reasoning was definitely something that… stood out.}''
Therefore, \kriya{}'s conversational interactions turned interpretation into a shared activity rather than a task users had to handle on their own. 
We also noted how participants built on \kriya{}'s offered explanations, to draw on their lived experience to 
confirm, refine, or push back on them. 
Because explanations were framed as possibilities rather than final answers, participants experienced interpretation as something they could actively participate in rather than passively receive. 
This shift reduced the burden of making sense of data in isolation. As P8 shared:


\begin{quote}
\small
``[\kriya{}] made it feel like something I didn't have to be afraid of, like not missing a goal or anything like that [..] It helped me understand that sometimes things are unplanned. So it's okay if you miss a goal for a couple of days.'' (P8)
\end{quote}

Together, these responses suggest that co-interpretation not only helped explain discrepancies, but also reshaped reflection as a collaborative reasoning process, setting the stage for participants to integrate contextual details and lived experience into how they understood wellbeing data.


\subsection{Compassionate Framing Shapes Willingness and Reflective Engagement}\label{results:emotional_framing}
Participants described that the tone and wording of \kriya{} played a role in their willingness to engage with their wellbeing data.
In this theme, we describe how non-judgmental language lowered the barrier to reflection and how flexible framing supported continued engagement when outcomes fell short. We elaborate this theme through two sub-themes: 1) \textit{Non-judgmental language lowers the barrier to reflection}, and 2) \textit{Flexible framing sustains reflection beyond performance}, as described below:

\subsubsection{Non-Judgmental language lowers the barrier to reflection}
Emotional tone in \kriya{}'s language of interaction played a central role in whether participants felt willing to engage with their data at all. 
Participants actually appreciated how the feedback was framed in an encouraging and explanatory way, and described 
feeling less emotionally guarded, especially when encountering values that might otherwise feel discouraging. 
This effect was particularly pronounced among P4, P9, P11, and P17 with little or no prior experience using wellbeing apps, who appreciated that the system did not assume familiarity with metrics or performance targets. 
For example, P17 described how \kriya{}'s linguistic framing shifted the experience away from evaluation, explaining that ``\textit{the evening debrief part is kind of courteous [..] it feels like it stays surprising and it is not judging},'' and instead felt ``\textit{more curious, like more open and prompting you to do better without judging why you have that.}'' 
Further, P11 expressed a similar experience, noting that the system felt ``\textit{in line with understanding what happened today instead of judging me for not just meeting my goals.}'' 
Overall, P4, P9, P11 and P17 appreciated that the language invited them to stay and reflect on their data, rather than provoking avoidance or self-criticism, or a \textit{judgment of success or failure}. 


\subsubsection{Flexible framing sustains reflection beyond performance}
We also found that \kriya{}'s flexible framing shaped how participants understood what reflection was for. 
Participants described the system as encouraging an ongoing, revisable process in which numbers served as signals of context, fluctuation, and potential adjustment. 
P7, P13, and P14 described feeling more comfortable when metrics were framed as ranges instead of strict pass/fail targets (e.g., through Comfort Zones). They also found explanations easier to accept when they connected deviations to an everyday context, since this could reduce the sense that a shortfall was a personal failure.

Participants mentioned how they may be demotivated by typical health apps which give a sense of success or failure, e.g., P16 mentioned, ``\textit{Even though I knew I had walked a fair distance that day, if I see the count as very small, it feels like I let myself down.}''
They mentioned that reflection with \kriya{} felt less about
evaluating whether a goal had been met and more about interpreting how daily circumstances shaped those outcomes. P14 shared this experience when describing the planning feature:

\begin{quote}
\small
``Actually, after planning, I feel motivated to achieve the plan. It has a positive effect, and I'm OK even if I didn't achieve the plan. But planning makes me put in a good way, a motivated way to achieve it.'' (P14)
\end{quote}

Similarly, P7 described this as feeling informative in a non-judgmental way, ``\textit{``It felt neutral. Not in like a bad way, but neutral, like non-judging. It was just helpful to get you the information you needed.''}


Overall, these perspectives highlight that emotional comfort came not only from reassurance but also from redefining success as continued sensemaking and adjustment, so reflection stayed worthwhile even when outcomes fell short.

\subsection{Factors for Trust, Understanding, and Continued Use}\label{results:trust_understanding_use}

We found that participants' interest in continuing to use \kriya{} depended on how they assessed its credibility, understood its explanations, and balanced interpretive depth with everyday effort. 
We also noted that participants' understanding and cognitive effort were intertwined with this trust calibration: interpretations felt valuable when they were legible and grounded, but could become burdensome when concepts or probabilistic outputs were difficult to parse. 
We elaborate on this through two sub-themes: 1) \textit{Trust through transparency and communicated uncertainties}, and 2) \textit{Tradeoffs between interpretive depth and cognitive effort}, as described below:

\subsubsection{Trust through transparency and communicated uncertainties}
We found that participants described trust in \kriya{} less in terms of whether its interpretations were always correct and more in terms of how the system communicated its reasoning. 
P11 and P13 expressed that trust increased when \kriya{} made clear that its interpretations were non-definitive, and when it explained why a particular conclusion was being suggested instead of presenting it as a settled fact.
Participants emphasized that seeing the system’s reasoning process helped them assess and contextualize its suggestions, rather than treating them as authoritative judgments.

Further, several participants noted that occasional inaccuracies were acceptable as long as uncertainty was made explicit. 
As P11 explained, the presence of visible reasoning and acknowledgment of uncertainty made the system feel more trustworthy, even when its interpretations did not fully align with their own experience: 

\begin{quote}
\small
``It’s fine if it’s not right all the time, as long as it’s kind of saying this is just a guess and not telling me this is definitely what happened.'' (P11)
\end{quote}

In this way, transparency around uncertainty reframed trust as something grounded in communicative openness rather than factual perfection.



Some participants (P6, P9, and P13) also evaluated credibility by checking whether \kriya{}'s explanations aligned with their own lived experience. 
When the system pointed to familiar patterns such as workload, schedule disruptions, or everyday constraints, its interpretations felt grounded and relatable rather than abstract. 
For example, P6 described recognizing themselves in the system’s reasoning, noting that it \textit{``actually made sense with what my day looked like, because I was sitting in meetings most of the time.''} 
This alignment helped participants treat explanations as plausible interpretations rather than speculative claims.

At the same time, participants were clear about where they wanted the system to stop inferring. 
Trust could diminish when errors distorted core signals or when explanations extended beyond available evidence. 
P5 described that ``\textit{if it messes up the steps that I made and interprets that in a wrong way, that gives a whole wrong picture of what’s happening.}'' 
P9 also expressed discomfort when the system ventured beyond available evidence, stating, ``\textit{It starts to feel weird when it asks things that I didn’t really give it information about.}'' 
Similarly, P8 described clear boundaries around the data they were willing to share: ``\textit{I'm fine with giving away things like my step data and my sleep hours [..] but if it started to get more personal, I might be a bit hesitant [..] I'm fine with where it is right now.}''

These responses suggest that credibility depended not only on whether explanations were provided, but on whether they remained appropriately bounded. Participants valued insight that clarified their data, but resisted interpretations that extended beyond what the data could reasonably support.

\subsubsection{Tradeoffs between interpretive depth and cognitive effort}
Participants' responses highlighted comprehension and cognitive effort as central factors shaping whether \kriya{} felt sustainable to use over time. 
While many appreciated the tool's interpretive depth, others described moments of confusion or mental overload, particularly when encountering constructs such as Comfort Zone, Surprise Score, and probability based forecasts. 
These reactions did not simply reflect usability breakdowns. Instead, they revealed a tension between supporting deeper reflection and preserving a low effort, quick check in experience that could fit into everyday routines.

P1 articulated this tension when encountering probabilistic outputs in confidence scores, asking, ``\textit{I'm a little bit confused on like the 30\%, like where are those percentages coming from?}'' 
Others emphasized a desire for lighter interactions in certain moments, with P8 explaining, ``\textit{I kind of want it to be like a simple click one thing and I'm done. This kind of feels like there's a lot going on}'' (P8).
Importantly, this tension was closely tied to when and how participants imagined using the system. As P6 noted:

\begin{quote}
\small
``I might do this in the morning, but it really depends. If it’s a busy day, I might not. In that case, the planning part would be more helpful.'' (P6)
\end{quote}

In fact, we noted that participants expressed a desire for flexibility in how much effort self-reflection required across different moments. 
Taken together, these accounts point to the importance of adaptive depth, allowing reflective systems to support both quick acknowledgment and deeper sensemaking in ways that align with users' routines, time constraints, and attentional capacity. 
\section{Discussion}
Our work explored a complementary direction for wellbeing technologies by shifting from monitoring and motivating toward compassionate conversations and co-interpretation. 
Rather than treating health metrics as targets to hit, our prototype, \kriya{} framed data as an invitation to ask ``why'' and to build explanations, emotional meaning, and low-stakes next-step possibilities together. 
We organize our discussion around recurring interview themes: co-interpretation, emotional experience, and trust.




\subsection{Rethinking personal informatics as a space for shared interpretation}\label{subsec:disc-interpretation}
Our findings in~\autoref{results:tracking_to_interpretation} revealed how participants described reflection as a process of explaining mismatches and integrating lived context, rather than as checking performance against a target. 
Personal informatics research describes reflection as moving from data capture to integration in everyday understanding and action~\cite{li2010stage,  epstein2015lived, cho2022reflection}. 
Our findings identified a long-standing limitation in this pipeline: interpretive labor. Even with precise metrics, users must still reconcile numbers with context, explain variability, and manage emotional responses to deviations must~\cite{kersten2017personal, pantzar2017living, epstein2015lived, cho2022reflection}. 

In our study, \kriya{} suggests a design move from ``showing data'' to ``sharing the work of explanation.'' 
We identified conversational interactions that operationalize this shift: prompting users about discrepancies to surface \textit{why} questions, framing explanations as hypotheses to keep them revisable, and inviting user-supplied context to validate, revise, or reject factors. 
These mechanisms make the interpretive work of engaging with personal health data both visible and collaborative, extending critiques that several tracking tools conflate reflection with monitoring or compliance~\cite{cho2022reflection, baumer2014reviewing, ahmadpour2017information}.

We also observed boundary conditions that clarify when shared interpretation breaks down. 
As we noted in \autoref{results:tracking_to_interpretation}, surprises were most helpful when they came with a clear explanation or a concrete next step. 
When that link was missing, participants described the experience as disconnected from what they were trying to do in the moment. 
Together, these findings specify how conversational co-interpretation can redistribute interpretive labor, and when it risks reverting to either "just numbers" or "too much going on."

\subsection{Designing for emotional safety without diluting insight}

\autoref{results:emotional_framing} revealed participants' willingness to engage depended on emotional framing, and this sensitivity varied by metric (e.g., sleep versus steps). Compassionate design emphasizes non-judgmental language, validation, and psychological safety~\cite{van2023role, ludden2024compassionate, lusi2025designing}. 
Our findings show that emotional experience is not a surface layer on top of analytics; it determines whether people feel able to engage with their data at all. 

Across interviews, participants noted that \kriya{}’s neutral and nonjudgmental language lowered the barrier to engaging with potentially discouraging data. Rather than interpreting low values as personal failure, they described feeling more open to considering what might have contributed to those outcomes. This effect was especially visible among participants with limited prior experience using wellbeing apps, who appreciated that the system did not assume familiarity with metrics or performance targets. In these cases, emotional framing helped participants remain engaged rather than avoidant when encountering unfavorable values.

At the same time, our results indicate that emotional comfort was closely tied to how explanations were presented. Supportive language was effective when it was paired with clear reasoning and concrete contextual factors, rather than generic reassurance \cite{kim2023help}. Participants were more receptive when explanations connected deviations to everyday circumstances or acknowledged uncertainty, which helped them treat feedback as informative rather than evaluative. 

Taken together, these findings extend compassionate design by showing that emotional attunement cannot be added post-hoc as a ``tone layer.'' Instead, it must be integrated into the analytic pipeline to examine what signals are surfaced, how uncertainty is framed, and how sensitive metrics are narrated, to prevent patronizing or judgmental experiences.

\subsection{Trust as a process: uncertainty, corrigibility, and boundary-respecting interaction}
\autoref{results:trust_understanding_use} revealed that participants calibrated trust through how \kriya{} communicated uncertainty, invited correction, and stayed within clear boundaries. 
Research around human-centered AI argues that appropriate reliance depends on intelligibility, user agency, and communication practices that calibrate trust~\cite{amershi2019guidelines, shen2022human,wang2019designing,miller2017explanation}. 
In the context of wellbeing reflection, our findings showed that trust emerged less from perceived accuracy, and more through interactional practices. 
Participants were more forgigiving to imperfect inferences when the system presented explanations as hypotheses, disclosed uncertainty, and invited correction. 

However, trust seemed to decrease when the system appeared to infer beyond available evidence or prompt sensitive claims without grounding (scope restraint). 
This provides context-specific evidence that corrigibility and scope restraint are not optional safeguards but core trust mechanisms in reflective wellbeing AI: users accepted "maybe" when it was explainable and revisable, but withdrew when the system overclaimed or exceeded its evidentiary scope~\cite{ribeiro2016trust}.
This low-stakes framing also appeared in participants’ descriptions of planning as exploratory rather than performance-driven when participants discussed motivation without pressure (\autoref{results:emotional_framing}). 
This is not gamification in the sense that there are no rewards, competition, streaks, or performance mechanics, which prior work links to guilt and pressure in self-tracking contexts~\cite{peterson2022self, lupton2016diverse}. 
Instead, playfulness functions as speculative sensemaking~\cite{elsden2017speculative, grossehering2013slow}. 
The What-If planning module illustrates how planning can be framed as curiosity when small alternatives become experiments to consider, not as commitments to meet. This expands the design space beyond dashboards and nudges toward dialogic exploration where the goal is understanding and agency, not compliance.

\subsection{Co-interpretation as complement, not replacement}
Although participants largely responded positively to KRIYA's conversational and compassionate framing, this does not necessarily imply that co-interpretive reflection is foolproof, and should replace existing goal-driven or performance-oriented wellbeing technologies. 
Prior work in personal informatics shows that dashboards, targets, reminders, and structured goals persist not by accident, but because they can be effective for many individuals and contexts—particularly when users seek accountability, measurable progress, or external structure~\cite{li2010stage, epstein2015lived, fitzpatrick2013review,sefidgar2024improving}. 
For some goals, especially those related to habit formation, training, or rehabilitation, explicit targets and performance feedback may provide productive friction that supports learning and change~\cite{consolvo2009theory}.

Our findings instead suggest that co-interpretive, curiosity-driven reflection represents a different mode of engagement, one that may be more suitable for certain moments, metrics, or individuals. 
Participants often described \kriya{} as lowering emotional barriers to engagement, particularly when they felt overwhelmed, inconsistent, or discouraged by traditional tracking tools. 
In this sense, KRIYA does not remove effort or challenge; rather, it reframes difficulty from failing to meet a standard to understanding why outcomes vary. This distinction echoes insights from learning and reflection-oriented systems, where psychological safety and interpretive space are understood as prerequisites for deeper engagement and sensemaking~\cite{ahmadpour2017information, morris2018towards}.

Importantly, participants' responses suggest that co-interpretive designs may appeal more strongly to individuals who are less motivated by strict goal-setting, who experience anxiety or disengagement around performance metrics, or who value reflection over optimization in their wellbeing practices. 
Prior work similarly highlights substantial individual differences in how people engage with self-tracking technologies and the meanings they attach to data~\cite{pantzar2017living, cho2022reflection}. 
For others---particularly those who already thrive under goal-driven regimes---\kriya{}'s exploratory framing may feel insufficient or overly indirect. 
Further, some individuals may actually prefer direct summative information rather than back-and-forth exchanges~\cite{duddu2025does}.
This points to an important implication: reflective AI companions should not be positioned as universal solutions, but as one dimension within a broader ecology of wellbeing tools~\cite{epstein2016beyond, kersten2017personal}.
Future work can further explore how co-interpretive, playful inquiry might be meaningfully integrated with goal-oriented approaches. 
For example, systems can allow users to move fluidly between exploratory reflection and structured goal-setting, or to selectively invoke compassionate interpretation during periods of stress, uncertainty, or disengagement. 
Understanding when users benefit from rigor versus when they benefit from inquiry---and how AI systems can adapt across these modes---remains an open direction for research in personal health informatics and human-centered AI~\cite{amershi2019guidelines,das2025ai,nepal2024mindscape,shen2022human}.

\subsection{Design implications for AI companions that support reflective wellbeing}

Overall, our findings suggest that tools like \kriya{} hold promise for enabling compassionate, collaborative, and emotionally attuned forms of wellbeing self-reflection. 
Importantly, this work points to several design implications for AI companions that support wellbeing reflection:

\para{Opportunity: Supporting Interpretive and Collaborative Reflection.} 
We started our work with the motivation that personal health dashboards typically present users with metrics and trends while offloading the work of explanation and interpretation onto individuals. 
Our findings suggest that AI companions have an opportunity to intervene at this gap. Rather than simply reporting outcomes, AI companions can help users interpret what those outcomes might mean in context.
These tools can prompt reflective "why" questions when patterns appear surprising, draw attention to mismatches between expectations and observed outcomes, and offer tentative explanations that users can agree with, revise, or reject~\cite{baumer2014reviewing, kersten2017personal, cho2022reflection}. 
Importantly, this interpretive support can help invite users to contribute their own context, such as schedule constraints, fatigue, or situational disruptions. 
When explanation becomes a shared activity, users are less likely to feel confused, blamed, or overwhelmed, and more likely to engage in reflection. 

\para{Replace binary goals with variability-aware expectations.} When framing wellbeing metrics around fixed targets or \textit{pass}--\textit{fail} thresholds, AI companions should represent expectations as flexible ranges that reflect everyday variability~\cite{peterson2022self, lupton2016diverse}. 
Participants responded more positively when fluctuations were shown more compassionately and normalized into contexts, especially when metrics were interpreted relative to sleep, workload, or routine changes. 
Probabilistic framings can further support this approach by communicating likelihood and uncertainty. 
Together, these strategies help users understand patterns without attaching moral weight to shortfalls, reducing pressure and supporting reflection that better matches the realities of daily life.

\para{Make emotional framing part of the analytic pipeline.} Emotional tone should not be layered above interpretation, but integrated into how interpretations are produced and communicated. Participants responded best when supportive language was clearly tied to reasoning, such as explaining which contextual factors may have contributed to an outcome, without it being offered as generic reassurance~\cite{van2023role, lusi2025designing, ludden2024compassionate, neff2011self}. 
Emotional sensitivity also differed by metric. 
Sleep data often carries greater emotional stakes than step counts, suggesting the need for metric-aware defaults that adjust tone, depth, and framing accordingly. 
Allowing users to influence how supportive or exploratory the system feels further respects individual preferences and emotional thresholds. 
When emotional framing is integrated into analytic decisions, AI companions can offer care along with clarity and encouragement.

\para{Build trust through corrigibility.} Trust in reflective AI systems should not depend solely on correctness. 
Alternatively, trust emerged when systems were open about uncertainty and allowed users to correct or refine interpretations. 
Explanations should be presented as hypotheses rather than directive conclusions, with visible cues about the source of information they derived from~\cite{amershi2019guidelines, shen2022human, wang2019theory, miller2017explanation, ribeiro2016trust}. 
Providing ways for users to push back, clarify context, or revise system assumptions helps position AI as a collaborator.



\para{Support contextual journaling and longitudinal summarization as shared memory.}
Participants often evaluated whether an explanation ``made sense'' by drawing on their own lived context, such as recent stressors, schedule disruptions, or atypical days. 
When these contextual details were absent, interpretations felt incomplete or unconvincing, placing the burden of sensemaking back on the individual. 
This suggests that wellbeing tools should support reflection not only through momentary insights but also by enabling lightweight contextual journaling---such as brief notes, tags, or calendar-linked entries---that can later be revisited. 
Participants also expressed interest in seeing simple longitudinal summaries that connect patterns to recurring situations while also acknowledging exceptions, rather than collapsing variability into a single performance narrative. 
Crucially, participants emphasized the importance of retaining uncertainty and control over what the system remembers and resurfaces, reinforcing prior calls for user-governed reflection~\cite{epstein2015lived, cho2022reflection}. Recent work on generative AI–powered journaling, such as MindScape~\cite{nepal2024mindscape} and DiaryMate \cite{kim2024diarymate}, offers complementary evidence that integrating contextual signals into longitudinal, AI-supported reflection can make journaling feel more personally meaningful and emotionally supportive over time.


\para{Make exploration playful while keeping scope boundaries clear.}
Our findings suggest that reflection can feel lighter when the system frames data as curiosity rather than a score. 
An AI companion can ask playful, memory-based questions grounded in a user's own records, then use the answer to start a short sensemaking conversation~\cite{elsden2017speculative}. It can also support ``what-if'' exploration as low-stakes experiments, with outputs labeled as hypotheses and shaped by user constraints~\cite{elsden2017speculative, grossehering2013slow}. When systems speculate, they should stay within clear evidence boundaries and communicate uncertainty, so users can calibrate trust and correct assumptions~\cite{amershi2019guidelines, wang2019theory, miller2017explanation, ribeiro2016trust}.
\subsection{Limitations and Future Directions}


Our study has limitations, which also suggest interesting directions for future research.
We used \kriya{} as a prototype~\cite{hutchinson2003technology} and relied on hypothetical scenarios and scripted interactions, not participants' own records and not live AI conversations. 
This decision lowered privacy risk and allowed consistent comparison across interviews. Yet hypothetical data can change the experience of reflection. When the numbers come from someone's real life, the stakes can feel higher, and people may respond with more concern or more resistance.
This also emerged as a criticism from participants who noted how \kriya{}'s chatbot abilities were currently limited, and it could only answer a limited set of questions.
Our findings also come from short, single-session walkthrough interviews. They do not show what happens after repeated use in everyday life. People's responses to self-tracking tools can change with time. Novelty can fade, errors can repeat, routines can shift, and some people may stop using the tool~\cite{clawson2015no, epstein2016beyond}. 
Therefore, our work does not claim behavior change, clinical outcomes, or forecasting accuracy.

These questions are best answered through longitudinal deployments in everyday settings. Such deployments with participants' own data can show how trust and comfort change over time. 
These studies also need clear consent and boundary controls, so participants can see what sources the system uses and what the system is allowed to infer~\cite{amershi2019guidelines, shen2022human}. 
A broader sample beyond college students can test whether the same interaction patterns hold in other age groups and health contexts. 
Self-tracking habits and emotional stakes can differ across cultures and life situations~\cite{lupton2016diverse, sharon2017self}. Our study did not examine how open-ended, multi-turn conversations unfold when the AI responds to a participant’s real-time disclosures and corrections. Conversational agents can also create risks that differ from dashboards, such as explanations that sound reasonable but are wrong. Future work can test how systems communicate uncertainty, how users correct them, and how systems avoid overstepping what the data supports~\cite{burr2018analysis, casu2024ai}. 
\section{Conclusion}

In this study, we designed \kriya{}, an AI companion prototype, that explores an alternative direction for wellbeing technologies: conversational co-interpretation of personal health metrics to help facilitate reflective sensemaking. 
Across semi-structured interviews with 18 participants using hypothetical scenarios, we found that \kriya{} encouraged participants to co-interpret and explain wellbeing data, compassionate framing shaped reflective engagement, and that trust was built through transparency, uncertainty signaling, and openness to user correction. 
Together, these findings expand the design space for AI companions in wellbeing by showing how systems can share interpretive labor, support self-compassion through contextual feedback, and enable low-stakes exploratory planning without reverting to goal-driven judgment. 




\bibliographystyle{ACM-Reference-Format}
\bibliography{0paper}


\appendix
\clearpage
\section{Appendix}\label{sec:appendix}
\setcounter{table}{0}
\setcounter{figure}{0}
\renewcommand{\thetable}{A\arabic{table}}
\renewcommand{\thefigure}{A\arabic{figure}}

\begin{figure*}[h!]
\begin{subfigure}[b]{\columnwidth}
    \centering
    \includegraphics[width=\columnwidth]{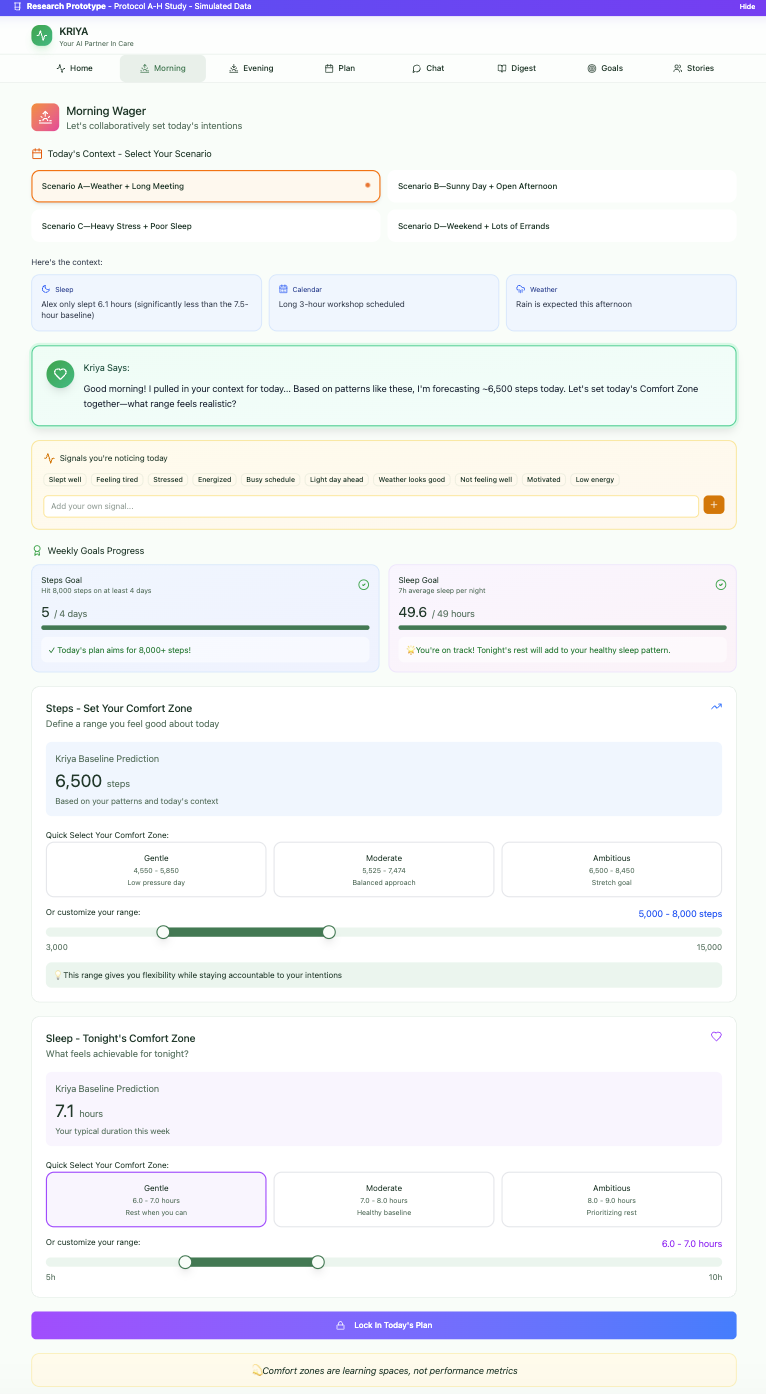}
    \caption{Morning}
    \end{subfigure}\hfill
  \begin{subfigure}[b]{\columnwidth}
    \centering
    \includegraphics[width=\columnwidth]{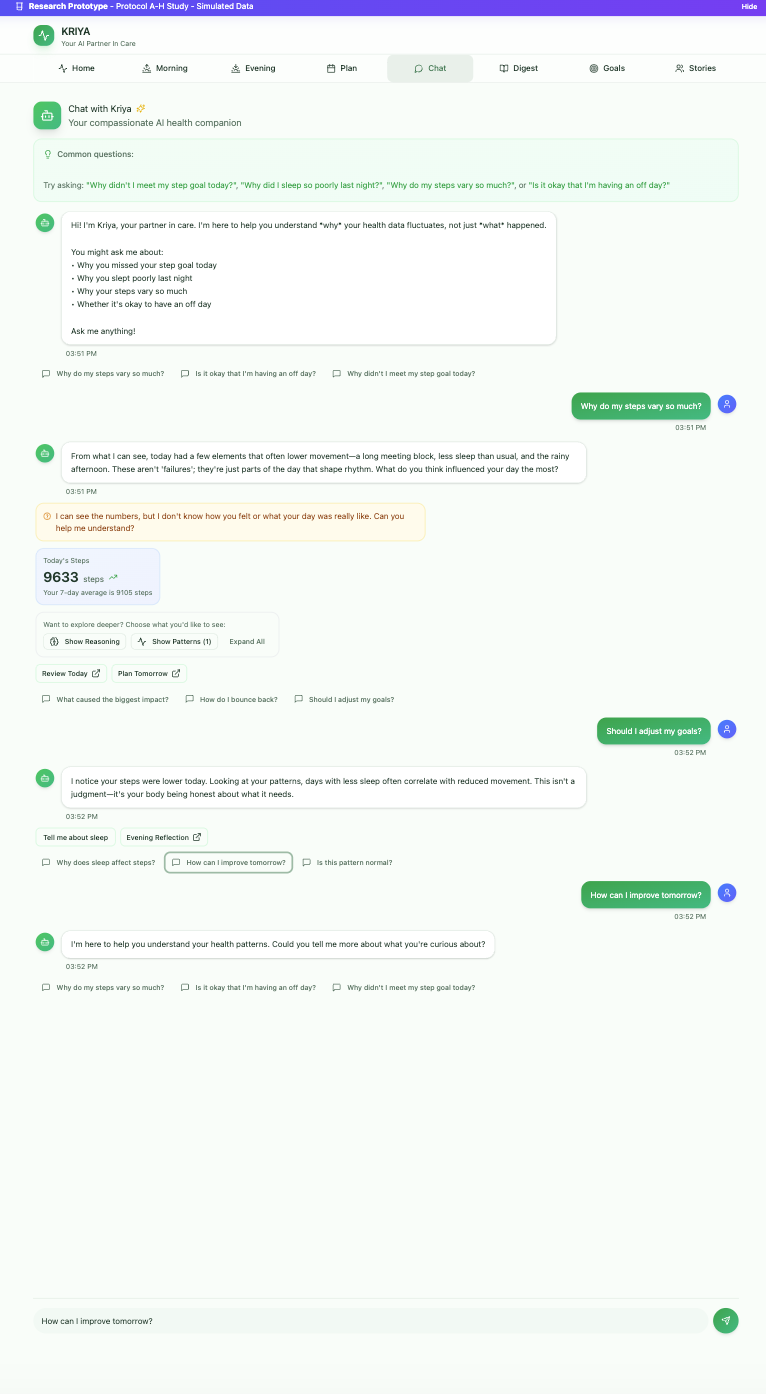}
    \caption{\kriya{} Chat}
    \end{subfigure}\hfill
    \caption{Screenshots of the \kriya{}. Modules include: (a) Morning Forecast where users set a Comfort Zone (a realistic range rather than a single goal) (b) a conversational interface for asking questions about steps and sleep.}
    \label{fig:kriya_morning}
    \Description[figure]{Two screenshots from the \kriya{} prototype. The left screenshot shows the Morning Forecast module, where the system provides a contextual overview for the day and prompts the user to choose a Comfort Zone, defined as a realistic range rather than a single goal. The right screenshot shows the \kriya{} chat interface, where the user can ask questions about steps and sleep and receive conversational responses that support interpretation and reflection.}
\end{figure*}

\clearpage
\begin{figure*}[t!]
 \begin{subfigure}[b]{\columnwidth}
    \centering
    \includegraphics[width=\columnwidth]{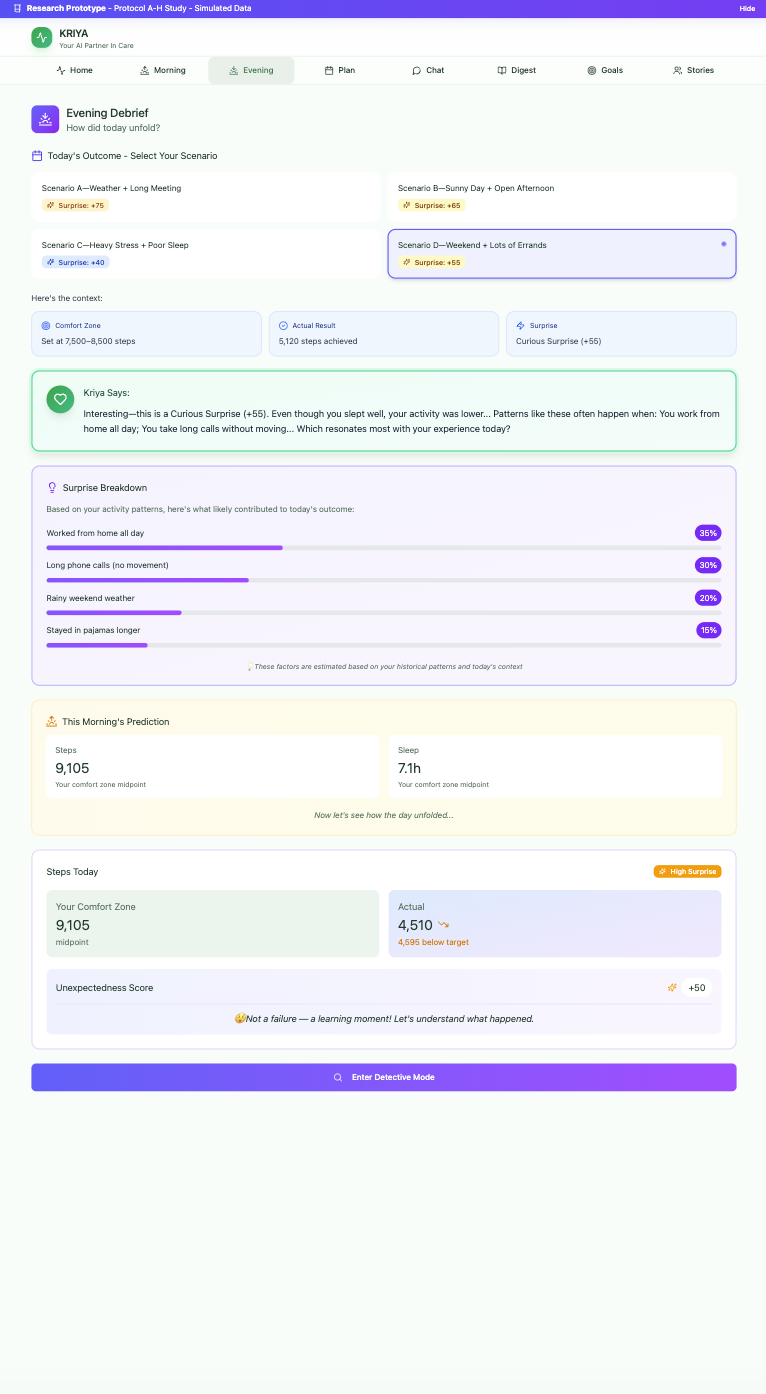}
    \end{subfigure}\hfill
\begin{subfigure}[b]{\columnwidth}
    \centering
    \includegraphics[width=\columnwidth]{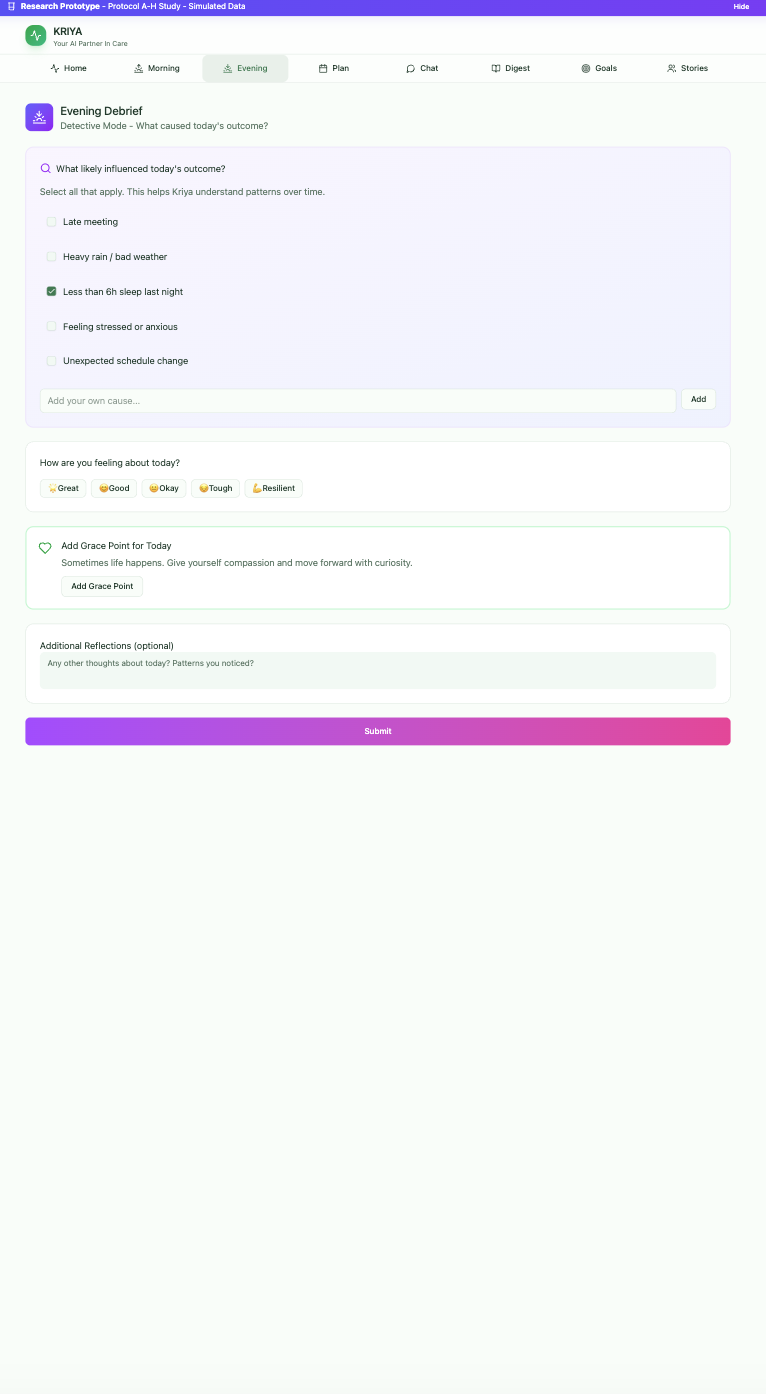}
    \end{subfigure}\hfill
    \caption{Screenshots of the \kriya{} prototype in Evening Debrief and Detective Mode that support explanation-oriented reflection.}
    \label{fig:kriya_evening}
    \Description[figure]{Two screenshots from \kriya{}’s Evening Debrief and Detective Mode. The screens support retrospective, explanation-oriented reflection by helping the user compare outcomes with earlier expectations (such as the selected Comfort Zone) and by presenting potential contributing factors to explain why the day’s results differed from what was anticipated.}
\end{figure*}

\end{document}

\endinput